\documentclass[11pt,a4paper]{amsart}

\usepackage{amsmath,amssymb,latexsym,amsthm,enumerate}
\usepackage{verbatim}
\usepackage{graphicx} 
\usepackage{hyperref} 
\usepackage{xcolor} 
\usepackage{psfrag} 
\usepackage{subfigure} 

\def\1{\mathbf{1}}
\def\0{\mathbf{0}}

\newcommand{\te}{\widehat{e}}

\newtheorem{prop}{Proposition}
\newtheorem{cor}{Corollary}

\def\BEN{\begin{enumerate}}  \def\BI{\begin{itemize}}
\def\EEN{\end{enumerate}}   \def\EI{\end{itemize}}
    \def\nn{\nonumber}

\def\mbb{\mathbb}

\def\te#1{\mathrm{e}^{#1}}

  \def\nn{\nonumber}   
     
 \def\q{\qquad}

\numberwithin{equation}{section}

\newenvironment{pr}{\vspace{1mm}\noindent\textbf{Proof.}}
                   {\vspace{-7mm}\begin{flushright}$\Box$\end{flushright}}

\usepackage[UKenglish]{babel}
\begin{document}

\title[Fast computation of vanilla options and implied volatilities]{Fast computation of vanilla prices in time-changed models
and implied volatilities using rational approximations}

\author{Martijn Pistorius}
\address{Department of Mathematics, Imperial College London}
\email{m.pistorius@imperial.ac.uk}
\author{Johannes Stolte}
\address{Department of Mathematics, Imperial College London}
\email{j.stolte09@imperial.ac.uk}
\date{\today\newline {{{\it Acknowledgements.}
We thank H\'{e}lyette Geman, Lane Hughston, Felicity Pearce, William Shaw, Mike Staunton
and the participants of the Fourth International
Conference on Mathematical Finance, South Africa,
for useful comments.}}}

\begin{abstract}
We present a new numerical method to price vanilla options quickly in time-changed Brownian motion models. The method is based on rational function approximations of the Black-Scholes formula. Detailed numerical results are given for a number of widely used models. In particular, we use the variance-gamma model, the CGMY model and the Heston model without correlation to illustrate our results.
Comparison to the standard fast Fourier transform method with respect to accuracy and speed appears to favour the
newly developed method in the cases considered. We present error estimates for the option prices. Additionally, we use this method to derive a procedure to compute, for a given set of arbitrage-free European call option prices, the corresponding Black-Scholes implied volatility surface. To achieve this, rational function approximations of the inverse of the Black-Scholes formula are used. We are thus able to work out implied volatilities more efficiently than one can by the use of other common methods. Error estimates are presented for a wide range of parameters.

\end{abstract}


\maketitle
\section{Introduction}
\noindent
The Black-Scholes model introduced in 1973 is arguably the best
known result in quantitative finance.
Four decades after its introduction, the model continues
to be widely used, especially as a universal benchmark model,
in part due to its tractability.
In a frictionless market
in which the asset price is modelled as a geometric
Brownian motion (GBM) with constant drift and constant volatility,
the price of a European put or call option
has a closed form: the celebrated Black-Scholes formula.
The Black-Scholes formula is employed by traders
to convert prices into units
of implied volatility. However, the presence of an implied volatility smile in
option markets contradicts the assumptions of the Black-Scholes model
and demonstrates that the returns are asymmetric and leptokurtic.
At shorter maturities the volatility smile becomes more pronounced,
showing an increasing deviation from the
GBM model. It is therefore well established that the GBM model
does not provide a satisfactory description of real market price dynamics.
Real market asset prices typically exhibit jumps and
volatility clustering,
while price paths in the Black-Scholes model are continuous
and have constant volatility.
These observations are of course well known: we refer the reader to
Cont \& Tankov~\cite{Cont2004} and Gatheral~\cite{Gatheral2006} for relevant background and further references.

To address the shortcomings of the GBM model,
a number of model classes with jumps or stochastic volatility
in the asset dynamics have been introduced.
These models can be broadly divided into
two categories: (a) jump-diffusion or finite activity models;
and (b) infinite activity models. In the first category, the evolution of prices is modelled by a jump-diffusion process, with
a finite number of jumps occurring at random times before a finite
time-horizon. The jumps can be interpreted as rare events,
crashes or large drawdowns. For models in this category, the jump part is typically given by a compound Poisson process. Examples of such models are
the Merton jump-diffusion model~\cite{Merton1976},
the Kou model~\cite{Kou2002} and the Bates model
with stochastic volatility~\cite{Bates1996}.
The second category comprises models with infinitely many
jumps in every time interval. It has been argued that such models give a more realistic description of the price process at various time scales
(see Madan~\cite{Madan2001}, Carr \textit{et al.}~\cite{Carr2002} and Geman~\cite{Geman2002}). These models are capable of generating
non-trivial small-time behaviour and, as shown in~\cite{Carr2002}, the Brownian component
is no longer needed.
Many of these models are {\it time-changed models} that
can be written as Brownian motions time-changed by an increasing
stochastic process. The idea of stochastic
time-changes was introduced into finance
by Clark~\cite{Clark1973}, who modelled the observed price
process as Brownian motion run on an independent
second process called the {\it clock}.

It is important for trading and risk management to be able (a) to price vanilla options in
models that provide an accurate description of market prices, and (b) to generate the
corresponding arbitrage-free volatility surface quickly and efficiently.
Fast calculation of vanilla option prices in the specified model
is also essential for carrying out the calibration of
the prices to market quotes in a timely fashion.
A financial institution active in
the derivatives market will seek to evaluate large
portfolios of vanilla options at speeds close to real time, in order to control
its risk positions.

This paper is devoted the development of a new method for the pricing of vanilla
options and the implied volatility surface in the class of
time-changed models, using rational approximations. Such approximations
are known to offer an efficient and accurate method for
computation of the
cumulative normal distribution function (see Abramowitz \&
Stegun~\cite{Abramowitz1965} for details), and could therefore
be expected to perform well for approximations of the
Black-Scholes formula, which is given in terms of
the cumulative normal distribution function. Rational function
approximations generally outperform polynomial approximations
in terms of computational efficiency, that is, for a given computational effort
smaller (global) errors can be achieved with rational function approximations
(see Morris~\cite{Morris1983}). For some functions, the
optimal rational function approximation is known further to be
able to achieve substantially higher accuracy than the
optimal polynomial approximation with the same number of coefficients.
Rational function approximations have been extensively
used in theoretical physics, engineering, statistics and economics.
For background on rational approximations and their use in economics,
refer to Ralston \& Rabinowitz~\cite{Ralston2001}~(Chapter~7).

The approach proposed here is based on rational function
approximation theory: drawing on this theory, an approximation for
the value of a vanilla option in a time-changed model
is constructed, taking the form of a linear combination of
a number of {\it negative exponential moments} of the clock.
This yields an explicit approximation for the value of a vanilla
option in those time-changed models for which
the Laplace transform (and hence any negative exponential moment)
of the clock is available in tractable form.
For many of the popular time-changed models
the Laplace transform of the clock is known in closed-form.

To provide a numerical illustration, the method was implemented for
the variance-gamma (VG) model, the CGMY model and the zero-correlation Heston model (the version of this model does not fall into the class of models studied here, but could be analysed by a two-dimensional extension, which is left for future research).
The rational function approximation method was found to yield
stable, fast and accurate results.
Comparison of the method to the standard fast Fourier transform (FFT) method (see e.g. Carr \& Madan~\cite{Carr19982})
with respect to speed
and accuracy, appears to be favourable for the rational function
approximation method in the cases considered. At this point it is worth mentioning that
several alternative and refined Fourier methods have recently been developed for the pricing
of European options, such as, for instance, the COS method of Fang~\&~Oosterlee~\cite{Fang2008} --- see Boyarchenko \& Levendorskii~\cite{Boyachenko2011} for an overview.
A detailed comparative study of all the different pricing methods is left for future research.



The rational function approximation method is then applied
to compute Black-Scholes implied volatilities.
%
The method uses the fact that the computation of the implied volatilities that correspond to
a given set of arbitrage-free call option prices can be reduced to the computation of
a single function of two variables. Contrary to Li~\cite{Li2008}, who proposes
to approximate the entire resulting surface by a two-dimensional approximation,
we perform a number of one-dimensional approximations and interpolate
between these for any given combination of input parameters. For a wide range of
parameters, the maximal approximation error of this approach
is bounded by $5.5 \times 10^{-5}$, refining to our required level of accuracy
the approximation of Li~\cite{Li2008}, which achieves a maximal error of
$ 3.3 \times 10^{-3}$ over the same region. It should be noted that
this procedure can be used independently of the results derived for option pricing and is not
limited to time-changed Brownian motion models. We refer the reader who is primarily interested in the
implied volatility computation to Section \ref{Sec:ImpliedVolatility}.

The remainder of the paper is organized as follows.
In Section \ref{Sec:LevyProcessInMathematicalFinance}
the theory for time-changed models is briefly reviewed.
In Section \ref{Sec:VanillaCallOption} the price of
an option in a time-changed Brownian motion model is expressed in
terms of a normalised Black-Scholes formula, and
the approximation to the option price is presented in Section \ref{Sec:RationalApproximations}.
Section \ref{Sec:ErrorEstimates} is devoted
to a study of the error of the method, the application
to computation of the implied volatility is given
in Section \ref{Sec:ImpliedVolatility}, numerical results
are reported in Section~\ref{Sec:NumericalResults}, and
concluding comments are given in Section \ref{Sec:Conclusion}.

\section{Time-Changed Brownian motion models for the asset dynamics}
\label{Sec:LevyProcessInMathematicalFinance}
\noindent
We assume frictionless markets and no arbitrage, and take as given
an equivalent martingale measure (EMM) $ \mathbb{Q} $
chosen by the market.
All stochastic processes defined in the following are assumed to
live on the complete filtered probability space
$ (\Omega, \mathcal{F}, \{ \mathcal{F}_t\}_{t \geq 0}, \mathbb{Q}) $.

We concentrate on the equity market and consider models in which the
stock price process $ (S_t)_{t\ge 0} $ under $ \mathbb{Q} $ can be represented as follows:
\begin{eqnarray}
S_t &=& S_0 e^{(r-q)t+X_t- \omega t}, \label{Eqn:StockPriceProcess} \\
X_t &=& \theta Z_t + \sigma W_{Z_t}, \label{Eqn:TimeChangedBM}
\end{eqnarray}
where $ ( W_t)_{t \geq 0} $ is a Brownian motion and $ (Z_t)_{t\ge 0} $ is an independent
stochastic process called the clock.

Throughout the paper, $ r \geq 0 $ and $ q \geq 0 $ denote the constant risk-free interest rate
and the constant dividend yield respectively, $ S_0 $ denotes the known stock price at time zero
and $S_t$ denotes the random stock price at time $ t $. The condition that $ (S_t e^{-(r-q)t})_{t\ge 0} $
is a martingale
will be guaranteed by an appropriate choice of the
mean-correcting compensator $\omega $ as follows:
\begin{eqnarray}
\omega = \frac{1}{t} \log \mathbb{E}[ e^{X_t}],
\label{Eqn:MeanCorrectingCompensator}
\end{eqnarray}
where $\mathbb{E}[ e^{X_t}]$ is assumed to be finite for all $t \ge 0$.
As can be seen, $ (X_t)_{t\ge 0} $ arises by subordinating a Brownian motion, with drift $ \theta $ and volatility $ \sigma $, by the subordinator $ (Z_t)_{t\ge 0} $.
Standard references on L\'{e}vy subordination include
Bertoin~\cite{Bertoin1996}~(Chapter~3) and Sato~\cite{Sato1999}. See
Cont \& Tankov~\cite{Cont2004} and Geman \textit{et al.}
\cite{Geman2001} for background on the role of subordination
in financial applications.

The clock $ (Z_t)_{t\ge 0} $ is required to be an increasing process
for which the Laplace transform at time $t$ should be available in
tractable form.
Typically, $ (Z_t)_{t\ge 0} $ is modelled by either a L\'{e}vy subordinator, as discussed in Section \ref{Sec:TimeChangingWithLevySubordinator},
or as a time integral of a positive diffusion, as discussed in Section \ref{Sec:TimeChangingWithTimeIntegral}.
Note that suitable models for the clock can also be constructed
by combining these two ingredients and therefore include
time-changes of the form $ Z_t = Z_t^1 + Z_t^2 $ or $ Z_t = Z_{Z_t^2}^1  $, where both $ (Z_t^1)_{t\ge 0} $ and
$ (Z_t^2)_{t\ge 0} $ can be (i) a L\'{e}vy subordinator,
(ii) an integral of
a non-negative process, or (iii) any other increasing process. It is worth noting
that the three stochastic volatility L\'{e}vy models
discussed by Carr \textit{et al.}~\cite{Carr2003} are of this form.
This follows since the models considered in~\cite{Carr2003}
are three L\'{e}vy models time-changed by a mean-reverting square root process,
where each of the L\'{e}vy models itself
(the normal inverse Gaussian model, the VG model and
the CGMY model) can be written as a time-changed Brownian motion.

\subsection{Time-changing with an independent L\'{e}vy subordinator}
\label{Sec:TimeChangingWithLevySubordinator}
Stochastic processes with independent stationary increments provide key examples of stochastic processes in continuous time and are used widely
in mathematical finance to model price dynamics.

A L\'{e}vy process $ (X_t)_{t \geq 0} $ has marginal distributions that
are infinitely divisible. Well-known examples of infinitely divisible laws are the Gaussian and the gamma distribution, which will
both be utilized in the remainder of this paper. Given a L\'{e}vy process $ (X_t)_{t \geq 0} $, define the corresponding characteristic function as follows:
\begin{eqnarray}
\Phi_{X_t}(z) \equiv \mathbb{E} [e^{izX_t}] = e^{-t \phi(z)}, \ \ z \in \mathbb{R},
\label{Eqn:characteristicfunction}
\end{eqnarray}
where $ \phi : \mathbb{R} \to \mathbb{R} $ is called the
characteristic exponent. The characteristic exponent can be shown
to exist and shown to
be a continuous function. By the L\'{e}vy-Khintchine representation,
it is also known that the characteristic exponent can be written as
\begin{eqnarray}
\phi(z) = \frac{1}{2} A^2z - i \gamma z + \int_{-\infty}^{\infty} \left( 1-e^{izx}+izx \mathbb{I}_{|x| \leq 1} \right)\nu(dx) ,
\end{eqnarray}
with characteristic triplet $ (A,\nu,\gamma)$.

A  L\'{e}vy subordinator $ ( Z_t)_{t \geq 0} $ is a L\'{e}vy process
that takes values in $ \mathbb{R_+} $ with characteristic triplet $ (0,\rho,\beta)$ satisfying $\rho((-\infty,0])=0$, $ \int_0^{\infty} (x \wedge 1) \rho(dx) < \infty $, and $ \beta \geq 0 $; that is, $ (Z_t)_{t \geq 0} $ has no diffusion component, only non-negative jumps of
finite variation, and a non-negative drift.
As a consequence, the trajectories of $Z$ are almost surely increasing. Since $ Z_t $ is a positive random variable for any positive $t$, it is natural
to describe its distribution using the Laplace transform:
\begin{eqnarray}
\mathbb{E} [e^{-u Z_t}] = e^{-t \psi(u)} = e^{-t \left( \beta u + \int_{0}^{\infty} \left( 1-e^{-ux} \right)\rho(dx) \right)}, \quad \quad \forall u \geq 0,
\label{Eqn:LaplaceTransform}
\end{eqnarray}
where
$ \psi(u) $ is called the Laplace exponent of $Z$. Since $ (Z_t)_{t \geq 0} $ is an increasing process, it can be interpreted as a ``time deformation'' and be used as a stochastic clock. Subordinating any L\'{e}vy process
(in particular, a Brownian motion) by another independent L\'{e}vy process will yield a new L\'{e}vy process (see Theorem 30.1 in Sato~\cite{Sato1999}).

We next present some explicit models in this class that
we will employ later.

\subsubsection{The variance-gamma model}
\label{Sec:VG}
\noindent
The VG process was first introduced to finance by Madan \& Seneta~\cite{Madan1990} in its symmetric version, and later extended to its asymmetric version by Madan, Carr \& Chang~\cite{Madan1998}. In the asymmetric version discussed here, the variance gamma process has three parameters: $ \theta \in \mathbb{R}$, $ \sigma > 0 $, $ \nu > 0$. It is defined by evaluating a Brownian motion with drift $ \theta $ and volatility $ \sigma $, as given in Equation (\ref{Eqn:TimeChangedBM}), at an independent gamma time. Specifically, the time-change $ (Z_t)_{t\ge 0} $ is now given by a gamma process independent of $ (W_t)_{t\ge 0} $ with marginal distribution at time $ t $ given by a gamma distribution $G(\frac{t}{\nu},\nu)$ with shape parameter $ \frac{t}{\nu} $ and scale parameter $ \nu $. The probability density function of $Z_t$, conditional on $ Z_0 = 0$, is given by
\begin{eqnarray}
f_{Z_t}(x) = \frac{x^{\frac{t}{\nu}-1} e^{- \frac{x}{\nu}}}{\nu^{\frac{t}{\nu}} \Gamma(\frac{t}{\nu})},
\label{Eqn:pdfVG}
\end{eqnarray}
and its Laplace transform has the form
\begin{eqnarray}
\mathbb{E} [e^{-u Z_t}] = (1 + \nu u )^{- \frac{t}{\nu}}.
\label{Eqn:LaplaceTransformVG}
\end{eqnarray}

\subsubsection{CGMY}
\noindent
The eponymous CGMY model was developed by Carr, Geman, Madan \& Yor~\cite{Carr2002} and can be seen as a generalization of the VG model discussed above. In particular, the class of tempered stable processes that can be represented as time-changed Brownian motion coincides with this model. The characteristic function of $X_t$ in this model can be shown to equal
\begin{eqnarray}
\mathbb{E} [e^{ i z X_t}] &=& e^{t C \Gamma(-Y) [ (M-i z)^Y+(G+i z)^Y - M^Y - G^Y]}.
\label{Eqn:CharFunCGMY}
\end{eqnarray}
The Laplace transform of $Z_t$, the value of the corresponding clock $ (Z_t)_{t\ge 0} $ at time $t$, is given by (see Madan \& Yor~\cite{Madan2008} for details)
\begin{eqnarray}
\mathbb{E} [e^{ - u Z_t}] &=& e^{tC \Gamma(-Y) [2v^Y cos(\eta Y) - M^Y - G^Y]}
\label{Eqn:LaplaceTransformCGMY} \\
v & = & \sqrt{2 u + GM} \nonumber \\
\eta & = & \text{arctan} \left( \frac{\sqrt{2 u - ((G-M)/2)^2}}{((G+M)/2)} \right) \nonumber.
\end{eqnarray}
For this model we set the drift equal to $ \theta = \frac{G-M}{2} $ and the volatility to $ \sigma = 1 $.

\subsubsection{Other examples} Another popular model that falls within
this category is the normal inverse Gaussian (NIG) model of Barndorff-Nielsen~\cite{Barndorff1998} and its extension to the generalized hyperbolic class by Eberlein, Keller \& Prause~\cite{Eberlein1998}.

\subsection{Time-changing with a time integral of a positive Markov process}
\label{Sec:TimeChangingWithTimeIntegral}
In this case the stochastic clock $(Z_t)_{t\ge 0}$ is defined as follows:
\begin{eqnarray}
Z_t = \int_0^t V_s ds,
\label{Eqn:ZtasIntegral}
\end{eqnarray}
where $ (V_t)_{t \geq 0} $ is a mean-reverting non-negative Markov process.
The mean-reversion is required to guarantee that the random time-change persists.

\subsubsection{Heston model without correlation}
\noindent
The Heston model~\cite{Heston1993} without correlation is obtained by defining $ (V_t)_{t \geq 0} $  in Equation (\ref{Eqn:ZtasIntegral})
to be equal to the positive CIR rate process introduced by Cox, Ingersoll \& Ross~\cite{Cox1985}. The stochastic differential equation for this square root process is given by
\begin{eqnarray}
d V_t = \kappa ( \delta - V_t) dt + \xi \sqrt{V_t} dB_t, \ \ \ \ \  t \geq 0,
\label{Eqn:StochasticDiffEquHeston}
\end{eqnarray}
where $(B_t)_{t\ge 0} $ is a standard Brownian motion that is independent of Brownian motion $ (W_t)_{t\ge 0} $ in Equation~(\ref{Eqn:TimeChangedBM}). Additionally, the CIR activity rate process starts at $ V_0 > 0 $, $ \kappa > 0 $ is the rate of mean reversion, $ \delta > 0 $ is the long-run activity rate level and $ \xi > 0 $ is the active rate volatility. It is further assumed that the Feller condition $ (2 \kappa \delta \geq \xi^2) $ is satisfied to ensure that the process never hits zero.

The characteristic function of $Z_t$ is well known from the work of Cox, Ingersoll \& Ross~\cite{Cox1985}, and is closely associated with L\'{e}vy's stochastic area formula. We recall that the Laplace transform of $ (Z_t) $ in this setting is given by
\begin{eqnarray}
\mathbb{E} [e^{ - u Z_t}] &=& A(t,u)e^{-B(t,u) V_0} \nonumber \\
&=& \left( \frac{2 \eta e^{\frac{(\eta + \kappa)t}{2}}}{(\eta+\kappa)(e^{\eta t} -1) + 2 \eta} \right)^{\frac{2 \kappa \delta}{\xi^2}} \text{exp} \left[ \frac{- 2 u (e^{\eta t} -1) V_0 }{(\eta+\kappa)(e^{\eta t} -1) + 2 \eta} \right], \label{Eqn:LaplaceTransformHeston}
\end{eqnarray}
where $ \eta = \sqrt{2 \xi^2 u + \kappa^2}$. In the setting of
\eqref{Eqn:StockPriceProcess}--\eqref{Eqn:TimeChangedBM}, the Heston model without correlation corresponds to
the parameter values $\sigma = 1$ and $\theta = -\frac{\sigma^2}{2} = -\frac{1}{2} $, so that
$ X_t = -\frac{1}{2} Z_t + W_{Z_t} $ and $\omega = 0$.

\subsubsection{Other examples} By taking $ (V_t)_{t \geq 0} $ to be an affine jump-diffusion process for example, the so-called
quadratic models (see Leippold \& Wu~\cite{Leippold2002}) arise.

\section{Vanilla Options in Time-Changed Models}
\label{Sec:VanillaCallOption}
\noindent
A European call option on a stock with maturity date $ T $ and strike price $ K $ is
a contingent claim that gives its holder the right, but not the obligation, to buy the stock at date $ T $ for a fixed price $ K $.
The arbitrage-free value of a call option at time zero can be expressed as the discounted conditional expectation of the payoff:
\begin{eqnarray}
C = e^{-rT} \mathbb{E}^{\mathbb{Q}} [(S_T - K)^{+}],
\label{Eqn:CallPayoff}
\end{eqnarray}
where the price of the stock $S_T$ at time $T$ is given in Equation~\eqref{Eqn:StockPriceProcess} and $ r \geq 0 $ denotes the constant risk-free interest rate as before.
 In the Black-Scholes model, the risk-neutral dynamics of $S$ are described by the exponential of a Brownian motion with drift,
\begin{eqnarray}
S_t = S_0 e^{(r-q-\frac{\sigma^2}{2})t+ \sigma W_t}, \label{Eqn:BSdynamics}
\end{eqnarray}
where $ \sigma \geq 0 $ is the volatility of the asset. The call price (\ref{Eqn:CallPayoff}) in this model is given by the Black-Scholes formula:
\begin{eqnarray}
C_{BS} (S_0,K,T,r,q,\sigma) &=& e^{-rT} \mathbb{E}^{\mathbb{Q}} [(S_0 e^{(r-q-\frac{\sigma^2}{2})T+ \sigma W_T} - K)^{+}] \label{Eqn:BSEquation} \\
&=& S_0 e^{-qT} N (d_1) - K e^{-rT} N(d_2), \nonumber \\
d_1 &=& \frac{\text{log}(S_0/K)+(r-q+\frac{\sigma^2}{2})T}{\sigma \sqrt{T}}, \nonumber \\
d_2 &=& \frac{\text{log}(S_0/K)+(r-q-\frac{\sigma^2}{2})T}{\sigma \sqrt{T}}, \nonumber
\end{eqnarray}
where $ N $ represents the standard normal cumulative distribution function.
The following parameterisation of the
 Black-Scholes formula will be considered in what follows:
\begin{eqnarray}
c_{BS}(v;x,\mu) = e^{\mu v} N \left( \frac{x}{\sqrt{v}}+ \left( \mu + \frac{1}{2} \right) \sqrt{v}\right)- e^{-x} N \left( \frac{x}{\sqrt{v}}+ \left( \mu - \frac{1}{2} \right) \sqrt{v} \right).
\label{Eqn:cBS}
\end{eqnarray}
This normalisation is similar to the non-dimensional form of the Black-Scholes formula proposed by Lipton~\cite{Lipton2001}~(Chapter~9.3).
The function $c_{BS}$ is related to the original
Black-Scholes formula as follows:
\begin{eqnarray}
C_{BS} (S_0,K,T,r,q,\sigma) &=& S_0 e^{-rT} c_{BS} \left( \sigma^2 T;\text{log}(S_0 / K),(r-q) / \sigma^2 \right).
\label{Eqn:RewrittenBS}
\end{eqnarray}
Typically, in time-changed models of the form given in
Equations~(\ref{Eqn:StockPriceProcess})--(\ref{Eqn:TimeChangedBM}),
no closed form formulae for call option prices are known.
The next proposition shows how the call price of the
time-changed model ($C_{TC}$) can be expressed
in terms of the normalised Black-Scholes formula in~\eqref{Eqn:cBS}.
\begin{prop}
\label{Prop:CTC}
If the process $ (S_t)_{t \geq 0} $ under $ \mathbb{Q} $ is given by
the system of Equations (\ref{Eqn:StockPriceProcess}) - (\ref{Eqn:TimeChangedBM}), where $ (W_t) $ and $ (Z_t) $ are independent,
the vanilla call option price with underlying $S$
can be expressed as
\begin{eqnarray}
C_{TC} (S_0,K,T,r,q,\theta,\sigma) &=&  S_0 e^{-(q+\omega)T} \mathbb{E}^{\mathbb{Q}}  \left[ c_{BS} \left( \sigma^2 Z_T; x_{TC}, \mu_{TC} \right)\right] \label{Eqn:CTC},
\end{eqnarray}
where
\begin{equation}\label{eq:mx}
\mu_{TC} = \frac{\theta}{\sigma^2} + \frac{1}{2}\q
\text{ and }\q
x_{TC} = \log \left( \frac{S_0}{K e^{(q-r+ \omega) T}} \right).
\end{equation}
\end{prop}
\noindent
The parameters $ S_0,K,T,r,q,\theta $ and $ \sigma $ are all constant and defined above. From Equation (\ref{Eqn:MeanCorrectingCompensator}) it follows that the mean-correction $ \omega $ is available in closed-form:
\begin{eqnarray}
\mathbb{E}[ e^{X_t}] = \mathbb{E} \left[ e^{\theta Z_t + \sigma W_{Z_t}} \right] = \mathbb{E} \left[ e^{ \left( \theta + \frac{\sigma^2}{2} \right) Z_t} \right] =
e^{\psi \left( \theta + \frac{\sigma^2}{2} \right) t} =  e^{\omega t},
\label{Eqn:MeanCorrectingCompensator2}
\end{eqnarray}
so that $ \omega $ is specified as follows:
\begin{eqnarray}
\omega = \psi \left( \theta + \frac{\sigma^2}{2} \right),
\label{Eqn:omega}
\end{eqnarray}
where $ \psi(s)=\frac{1}{t}\log E[\te{s Z_t}] $ is the Laplace exponent  of $(Z_t)_{t\ge 0}$. In particular, it should be noted that $ \omega $ only changes with $ \theta $ and $ \sigma $ and stays constant once these market parameters have been specified. In the Heston model
with zero correlation, $ \omega $  is equal to zero, which follows
from the fact that $e^{X_t} $ is
a true martingale in the Heston model.

{\bf Remark.}
Note that $ \mu_{TC} $ only depends on the {\em market} parameters $\theta$ and $\sigma^2$ while $x_{TC}$ depends on $ r $, $ q $, $ S_0 $, $ T $, $ K$, $ \theta $, $ \sigma $ and $ \omega $. In particular, $ x_{TC} $ is {\em contract}-dependent
as it varies with maturity $T$ and strike $K$. We will refer to $ \mu_{TC} $ and $ x_{TC} $ as the
market-dependent parameter and the adjusted log-moneyness, respectively.

\begin{pr} By conditioning on $Z_T$ and given that $ (Z_t)_{t\ge 0} $ and $(W_t)_{t\ge 0}$
are independent, we find the following:
\begin{eqnarray*}
C_{TC} (S_0,K,T,r,q,\theta,\sigma) &=& e^{-rT} \mathbb{E} [(S_T - K)^{+} ] = e^{-rT} \mathbb{E} \left[ \left( S_0 e^{(r-q)T + X_T - \omega T} - K \right)^{+} \right] \\
&=& e^{-rT} \mathbb{E} \left[ \left( S_0 e^{(r-q) T+ \theta Z_T + \sigma W_{Z_T} - \omega T} - K \right)^{+} \right] \\
&=& e^{-(q+\omega)T} \mathbb{E} \left[ \left( S_0 e^{\theta Z_T +\sigma W_{Z_T}} - K e^{(q-r+\omega)T} \right)^{+} \right] \\
&=& e^{-(q+\omega)T} \mathbb{E} \left[ \left(S_0 e^{ (\mu_{TC} \sigma^2 -\frac{\sigma^2}{2}) Z_T + \sigma W_{Z_T}} - K e^{(q-r+\omega)T} \right)^{+} \right] \\
&=& e^{-(q+\omega)T} \mathbb{E} \left[ e^{\mu_{TC} \sigma^2 Z_T } C_{BS}(S_0, K e^{(q-r+\omega)T} ,Z_T,\mu_{TC} \sigma^2,0,\sigma) \right] \\
&=& S_0 e^{-(q+\omega)T} \mathbb{E} \left[ c_{BS} \left( \sigma^2 Z_T, x_{TC}, \mu_{TC} \right) \right],
\end{eqnarray*}
where we used the definitions \eqref{Eqn:BSEquation} and \eqref{Eqn:RewrittenBS} of $C_{BS}$  and $c_{BS}$ . \end{pr}

In the case that $(e^{X_t})_{t\ge 0}$ is a martingale (or equivalently,
$\theta=-\frac{\sigma^2}{2}$),
the normalised Black-Scholes formula $c_{BS}$ admits a symmetry:

\begin{cor}
\label{Cor:SymmetrycBS}
If $ \mu = 0 $, then the following holds true for all $v\in\mbb R_+$
and $x\in\mathbb R$:
\begin{eqnarray}
c_{BS}(v;-x,0) &=& 1-e^{-x}+e^{-x}c_{BS}(v;x,0). \label{Eqn:SymmetryIncTC}
\end{eqnarray}
\end{cor}
\noindent
This identity thus applies, for example, to the Heston model with
zero correlation.

\begin{pr}
By straightforward algebra we get the following equalities:
\begin{eqnarray*}
c_{BS}(v;-x,0) &=& N \left( \frac{-x}{\sqrt{v}}+\frac{\sqrt{v}}{2} \right) - e^{-x} N \left( \frac{-x}{\sqrt{v}}-\frac{\sqrt{v }}{2}\right) \\
&=& \left[1 - N \left( \frac{x}{\sqrt{v}}-\frac{\sqrt{v}}{2} \right) \right] - e^{-x} \left[1 - N \left(\frac{x}{\sqrt{v}}+\frac{\sqrt{v}}{2}\right) \right] \\
&=& 1-e^{-x} + e^{-x} \left[ N \left(\frac{x}{\sqrt{v}}+\frac{\sqrt{v}}{2}\right) - e^{x} N \left(\frac{x}{\sqrt{v}}-\frac{\sqrt{v}}{2} \right) \right] \\
&=& 1-e^{-x} + e^{-x} c_{BS}(v;x,0).
\end{eqnarray*}
\end{pr}

The discussion so far has only considered call options. However, with the help of the put-call parity, one can easily obtain
the value of a put option if the value of the corresponding call option is known. For convenience, we state the put-call parity
for the time-changed setting considered here:
\begin{eqnarray}
P_{TC} (S_0,K,T,r,q,\theta,\sigma) = C_{TC} (S_0,K,T,r,q,\theta,\sigma) + K e^{-r T} - S_0 e^{-q T}.
\label{Eqn:PCParity}
\end{eqnarray}

On account of Proposition \ref{Prop:CTC}, the problem of
pricing a call option in a time-changed model reduces to the problem
of  evaluating $ \mathbb{E} [ c_{BS} (\sigma^2 Z_T;x,\mu)] $
for known values of $ x $ and $ \mu $, which we address in the following section.

\section{Rational Approximations}
\label{Sec:RationalApproximations}
\noindent
To evaluate $C_{TC}$ given in Equation (\ref{Eqn:CTC}) we approximate
the function $v\mapsto c_{BS}(v; x, \mu)$ on a specified range $ v \in [a,b]$ by a rational function,
for given values of $ x $ and $ \mu $. For an integer $ m $, we denote the rational function approximation
of $c_{BS}(v; x, \mu)$ with degree $ m $ by $ c_{RA}^m (v; x, \mu) $:

\begin{eqnarray}\label{Eqn:RAc}
c_{BS}(v; x, \mu) \approx c_{RA}^m (v; x, \mu)
= \frac{\sum_{j=0}^m a_j^{x,\mu} v^j}{\sum_{j=0}^m b_j^{x,\mu}  v^j}.\\
\nn
\end{eqnarray}

\noindent To obtain the parameters $ a_j^{x,\mu}  $ and $ b_j^{x,\mu}$,
for given $ x $ and $ \mu$, we use a rational Chebyshev approximation
(see Appendix \ref{Sec:App:RationalApproximations} for a summary).

\medskip

Next, using a partial fraction decomposition, we can rewrite the right-hand side of Equation~(\ref{Eqn:RAc}) as follows:
\begin{eqnarray}
c_{RA}^m (v; x, \mu) =  A^{x,\mu}_0 + \sum_{j=1}^m \frac{A_j^{x,\mu} }{v-B_j^{x,\mu} }.
\label{Eqn:PartialFractionRA}
\end{eqnarray}
The parameters $ A_j^{x,\mu} $ and $ B_j^{x,\mu} $ will usually be
complex numbers.
If $Re(B_j)<0$, which will typically be the case
(see also the remark below), we may rewrite any of the denominators $(v-B_j^{x,\mu})^{-1}$ in integral form as
\begin{eqnarray*}
\frac{1}{v-B_j^{x,\mu}} = \int_0^\infty e^{-(v-B_j^{x,\mu} ) y} dy.
\end{eqnarray*}
Interchanging sum and integration yields the following:

\begin{eqnarray}
c_{RA}^m (v;x,\mu) = A_0^{x,\mu} + \int_0^{\infty} \left( \sum_{j=1}^m A_j^{x,\mu} e^{B_j^{x,\mu} y} \right) e^{- v y} dy.
\label{Eqn:InfiniteIntegralRA2}
\end{eqnarray}

\medskip

To approximate the integral in Equation (\ref{Eqn:InfiniteIntegralRA2})
efficiently, we truncate the integral and use a Gaussian quadrature:
\begin{eqnarray}
\int_{0}^{\infty} f(x) dx \approx \int_{c}^{d} f(x) dx
\approx \sum_{k=1}^L w_k f(x_k), \label{Eqn:GaussianQuad}
\end{eqnarray}
where $0\leq c< d<\infty$,  $w_k$ are the quadrature
weights, $x_k$ the abscissas and $f$ denotes
the integrand in \eqref{Eqn:InfiniteIntegralRA2}.
We compared a number of different Gaussian quadrature methods and
found that Gauss-Legendre performs particularly well; it
results in small errors for all parameter values
and models considered here. Abscissas and weights are
calculated by the standard procedure
(see Press \textit{et al.}~\cite{Press2002} for details)
and do not depend on the integrand $f$
in the case of Gauss-Legendre quadrature.
We thus obtain the following approximation:
\begin{eqnarray}\label{Eqn:cRAgaussianQuad}
\ \ \ c_{RA}^m (v;x,\mu) \approx A_0^{x,\mu} + \sum_{k=1}^L \left( w_k \sum_{j=1}^m A_j^{x,\mu} e^{B_j^{x,\mu} x_k} \right) e^{- v x_k}, \ \ \ \text{if} \ \ Re(B_j^{x,\mu}) < 0 \ \ \  \forall j.
%
\end{eqnarray}
To summarize, given constant values for $ x $ and $ \mu $, $ \mathbb{E}^{\mathbb{Q}} [ c_{BS} (\sigma^2 Z_T;x,\mu)] $ can be approximated by a rational function of the form
\begin{eqnarray}
\label{Eq:CTC2}
&&\mathbb{E}^{\mathbb{Q}} [ c_{BS} (\sigma^2 Z_T;x,\mu)]  \approx A_0^{x,\mu} + \sum_{k=1}^L \left( w_k \sum_{j=1}^m A_j^{x,\mu} e^{B_j^{x,\mu} x_k} \right) \mathbb{E}^{\mathbb{Q}} \left[ e^{- \sigma^2 Z_T x_k}\right], \  \text{if } Re(B_j^{x,\mu})<0 \  \forall j,
\end{eqnarray}
where $ A_j^{x,\mu} $ and $ B_j^{x,\mu} $ are the parameters of a rational function approximation with degree $ m $ (after applying a partial fraction decomposition), all depending on $ x $ and $ \mu $. Indeed, if $ Re(B_j^{x,\mu})<0$ for all $ j $,
it follows from Equation (\ref{Eqn:RAc}) and Equation~(\ref{Eqn:cRAgaussianQuad})
that the following approximation is valid:
\begin{eqnarray*}
c_{BS} (v; x, \mu) \approx A_0^{x,\mu} + \sum_{k=1}^L \left( w_k \sum_{j=1}^m A_j^{x,\mu} e^{B_j^{x,\mu} x_k} \right) e^{- v x_k}.
\end{eqnarray*}
Substituting $v=\sigma^2 Z_T$ and taking expectations on both sides yields the following expression:
\begin{eqnarray*}
\mathbb{E}^{\mbb Q} \left[ c_{BS} (\sigma^2 Z_T; x, \mu) \right] &\approx& \mathbb{E}^{\mbb Q} \left[ A_0^{x,\mu} + \sum_{k=1}^L \left( w_k \sum_{j=1}^m A_j^{x,\mu} e^{B_j^{x,\mu} x_k} \right) e^{- \sigma^2 Z_T x_k} \right],
\end{eqnarray*}
which is equal to the right-hand side of \eqref{Eq:CTC2}, by the linearity of the expectation.

Combining Equation~\eqref{Eq:CTC2} with Proposition \ref{Prop:CTC} leads to
an explicit approximation of the call option prices in the time-changed model
in terms of the Laplace transform of the stochastic clock $ Z_T $.

{\bf Remark.} Typically, the values $B_j^{x,\mu}$
in the rational approximation have negative real part, but
sometimes a coefficient $B_j^{x,\mu}$ has non-negative real part.
If this is the case, the easiest way to proceed is to increase
the order of the rational function approximation.
In the cases where switching to a higher
order approximation does not resolve the problem, one can deal with
the approximation of terms for which $Re(B_j^{x,\mu})\ge 0$ separately.
For those terms, we define a new variable $ v^* = e^{-v} $ and approximate the resulting function by a Chebyshev approximation of order $n$ to get
\begin{eqnarray}
\label{Eqn:CorrectionTerm}
&&\ \ \mathbb{E} \left[\frac{A_j^{x,\mu}}{v-B_j^{x,\mu}}\right] = \mathbb{E} \left[ \frac{-A_j^{x,\mu}}{\text{log}(v^*)+B_j^{x,\mu}}\right] \approx \mathbb{E} \left[\sum_{i=0}^n c_i^{x,\mu} (v^*)^i \right] = \mathbb{E} \left[ \sum_{i=0}^n c_i^{x,\mu} e^{- i v} \right] = \sum_{i=0}^n c_i^{x,\mu} \mathbb{E} \left[ e^{- i v} \right]
\end{eqnarray}
(for a review of Chebyshev approximations see Appendix \ref{Sec:App:ChebyshevApproximations}). Note that if more terms with $Re(B_j^{x,\mu})\ge 0$ occur,
only a single Chebyshev approximation is needed for all these terms
simultaneously. Thus, in Equation (\ref{Eq:CTC2})
the terms in the sum corresponding to $ B_j^{x,\mu} $ with non-negative
real part should be replaced by an approximation of the form given in Equation \ref{Eqn:CorrectionTerm}.

{\bf Remark.} In the case of the Heston model with correlation, a similar conditioning approach would lead to a family of functions of two variables that needs to be approximated. This follows since $ \text{log} \left( S_t / S_0 \right) $ is normally distributed conditional on $V_t$ and $ \int_0^t V_s ds $ (a fact which can be found in Glasserman~\&~Kim~\cite{Glasserman2009} or Broadie~\&~Kaya~\cite{Broadie2006}).
For given values $ \mu $ and $ x $, one would therefore approximate a two-dimensional function, which can then be rewritten in terms of negative exponential moments of the clock. We did so utilizing the downhill simplex search method
and results were reasonable. However, a detailed analysis of this case is left for future research, where one should investigate faster and more powerful two-dimensional approximation methods.

\subsection{Example: Variance Gamma model}\label{exam:raVGintegral}
This example illustrates how to price a vanilla call option in the VG model with the following parameters  (taken from the original paper by Madan \textit{et al.}~\cite{Madan1998})
$$
\sigma = 0.1213, \q  \nu = 0.1686\q \text{and}\q  \theta = -0.1436.
$$
We assume that $ S_0 = 1 $, $ K = 1.1 $, $ r = 0.03$, $ q = 0.01 $ and $ T = 1 $, so that we are pricing an out of the money call option with one-year maturity. It then follows from Equation \eqref{eq:mx}
that  $\mu_{TC}$ and $x_{TC}$ take the following rounded values:
$$ \mu_{TC} = -9.2596, \qquad   x_{TC} = 0.0594. $$
A rational approximation of $  c_{BS}(v; x, \mu) $ with degree $m=5$
is then given as follows:
\begin{eqnarray*}
c_{RA}^5 (v; 0.0594 , -9.2596) &=& -0.0041 - \frac{0.0009 + 0.0021i}{v+(0.0387-0.0667i)} - \frac{0.0009 - 0.0021i}{v+(0.0387+0.0667i)} \\
&+& \frac{0.0034}{v+0.0356} + \frac{8.0 \times 10^{-5}}{v+0.0098} + \frac{2.5 \times 10^{-5}}{v+0.0014},
\end{eqnarray*}
for $ v \in [0.0027,0.0405] $ after applying the partial fraction decomposition (it should be noted that we rounded these values only to present them here, but use more accurate values to derive the results stated below). Then, if we denote each ratio by $ \frac{A_j}{v-B_j} $, using the gamma distribution, the above can be rewritten as
\begin{eqnarray*}
c_{RA}^5 (v; 0.0594 , -9.2596) &=&  A_0 + \int_0^{\infty} \left( A_1 e^{B_1y} + A_2 e^{B_2y} + A_3 e^{B_3y} + A_4 e^{B_4y} + A_5 e^{B_5y} \right) e^{-y v} dy,
\end{eqnarray*}
where all five $ B_j $s have negative real parts.

\begin{figure}[t]
\scalebox{0.5}{\includegraphics{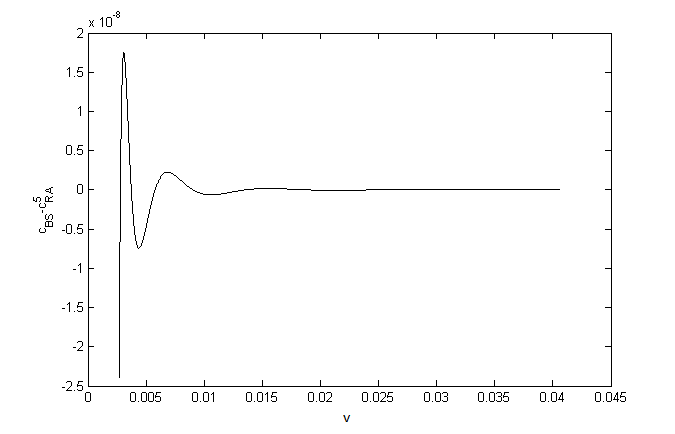}}
\caption{\small Plotted is the difference between the normalised Black-Scholes formula, $ c_{BS}(v; 0.0594 , -9.2596) $
given in \eqref{Eqn:RAc}, and the corresponding rational approximation with degree~5, $ c_{RA}^5 (v; 0.0594 , -9.2596) $, for $ v \in [0.0027,0.0405] $ with spot $ S_0 = 1 $, strike $ K = 1.1 $, interest rate $ r = 0.03$, dividend yield $ q = 0.01 $ and
maturity $ T = 1 $. \label{fig:MaxErrorExample1} }
\end{figure}

In Figure \ref{fig:MaxErrorExample1}, the difference of $  c_{BS}(v; 0.0594 , -9.2596) $ and $ c_{RA}^5 (v; 0.0594 , -9.2596) $ is plotted
for $ v \in [0.0027,0.0405]$.
The maximum error for this example is $2.39 \times 10^{-8} $.
Finally we use Gaussian quadrature to approximate
the infinite integral. Employing Proposition \ref{Prop:CTC} and Equation (\ref{Eq:CTC2}) to calculate the option price
yielded $ C_{TC} = 0.021403241$. By way of comparison, we also computed the value using
Carr-Madan \cite{Carr19982}'s FFT algorithm with the recommended values of the
dampening parameter ($\alpha=1.5$), step-size ($\eta=0.25$) and the number of steps ($N=2048$)
which yielded the value $0.021403243$, which differs from $C_{TC}$ by $2\cdot 10^{-9}$.

\subsection{Offline calculation and interpolation}
\label{Sec:splining}
\noindent
If one is interested in pricing a number of options with different
strikes and maturities, some practical improvements that yield substantial gains
in computation time can be made.
Given the market parameters, and therefore $ \mu_{TC} $, one
approximates the function $ c_{BS} (v; x, \mu_{TC}) $ for a number of values
of the adjusted log-moneyness $ x $ between $x_{TC}^{min} $
and $x_{TC}^{max} $ (30 values of $ x $ should usually suffice).
To calculate $x_{TC}^{min} $ and $x_{TC}^{max} $, one only needs
the smallest and largest strike and maturity that one is interested
in pricing, as all other values of $ x_{TC} $ lie between these
four corners. One then stores the corresponding approximation parameters
$ A_j^{x,\mu} $ and $ B_j^{x,\mu} $ for each value of $ x $ and uses an interpolation method (cubic splining, for example) for values in between.
As a consequence, one is able to speed up the method dramatically,
since any option price with these market parameters can now be calculated only by using interpolation.
Although the rational Chebyshev approximations can be performed very quickly, it should only be done once for a given
set of market parameters.\\
Obviously, when approximating only a few values of $ x $ offline, which are then used for the interpolation,
one should make sure that the errors of these approximations are sufficiently small.
As illustrated in the next section, for some of the $ x $ values, errors are
much bigger than for others, due to numerical instability.
However, each approximation can be carefully checked by evaluating the resulting errors (see Section \ref{Sec:ErrorOfRationalFunctionApproximation}),
as the computational time for these approximations should not matter as much (given that it is done only once offline).
One can therefore search for a rational function approximation that performs well for the given value of $ x $.

{\bf Remark.} When using this method for
fast calibration, one could pre-calculate approximations of
$ c_{BS} (v; x, \mu) $ for a number of parameter combinations of
$ x $ and $ \mu $.
The online calculation of an option price would then involve
a two-dimensional interpolation between the values of
$ x $ and $ \mu $ calculated offline. This should be particularly
feasible for calibrations where parameters are not allowed
to vary too much from parameters obtained in the previous
calibration, as one then has a well-defined range for $ \mu $.
Investigation of this idea is left for future research.

\section{Numerical Error Estimates}
\label{Sec:ErrorEstimates}
\noindent
Observe that the error from our method arises from the approximation
in Equation (\ref{Eq:CTC2}) only. This error can be subdivided into: (a) truncation error of $ v $; (b) error in the rational function approximation; and (c) error
in the Gaussian quadrature. In attempts to control
the error, there will be a
trade-off between the different sources of
error (a) and (b), that is the size of the truncation interval for $ v $ and the quality of the rational approximation
for a given degree. Since $ c_{BS} (v; x, \mu) $ decays as $ v^{1/2} $
for small $v$, approximating
this function accurately for very small values of $ a $ requires a high
order of the rational approximation. A suitable choice of the truncation limits $ a $ and $ b $
is therefore required to balance these two different sources of error.
In order to balance these errors and to find suitable choices for $ a $ and $ b $,
we analyse the cumulative distribution function (cdf) of $ \sigma^2 Z_T $ in Section \ref{Sec:pdfofM}.
Later, we consider the resulting error of the rational function approximation for the range
$ [a,b] $ and outside the range $ [a,b] $ in Section \ref{Sec:ErrorOfRationalFunctionApproximation}.

The third source of error is introduced by the numerical integration of the infinite integral in Equation (\ref{Eqn:InfiniteIntegralRA2}).
The numerical integration method should be chosen such
that the resulting error is comparable to the other two errors. If necessary, an adaptive quadrature method can be used to control for
the error of this numerical integration.

In order to further analyse the error that arises from the method presented here, we have chosen five sets of market parameters from the literature. We consider options with strikes $ K = 0.8, 0.81, 0.82, ..., 1.19, 1.2 $ and maturities $ T = 0.25, 0.5, 1, 1.5, 2, 2.5 $ for each of the five parameter sets detailed below,
and set $ S_0 = 1, r = 0.03 $ and $ q = 0.01 $. This leads to a total of $ 246 $ options and therefore $ 246 $ different values of $ x_{TC} $ for each set of market parameters. Table \ref{Tab:ChosenParameters} summarizes the five cases and states the source of each parameter set. Values of $ \mu_{TC} $, $ x_{TC}^{max} $ and $ x_{TC}^{min} $ are also given, where  $ x_{TC}^{max} $ and $ x_{TC}^{min} $ represent the maximum and the minimum of the $ 246 $ values of
the adjusted log moneyness $ x_{TC} $ for each case.
\tiny
\begin{table}[ht!]
\begin{center}
\begin{tabular}{llllrrr}
\hline
      Case &      Model & Parameters &     Source & $\mu_{TC} $ & $x_{TC}^{max} $ & $x_{TC}^{min} $ \\
\hline
         I &         VG & $ \sigma = 0.1213, \nu = 0.1686, \theta = -0.1436 $ & Madan \textit{et al.}~\cite{Madan1998} &    -9.2596 &     0.6099 &    -0.1436 \\

        II &         VG & $ \sigma = 0.178753, \nu = 0.13317, \theta = -0.30649 $ & Fiorani~\cite{Fiorani2003} &    -9.0920 &     0.9857 &    -0.1061 \\

       III &     Heston & $ \kappa = 0.87, \delta = 0.07, \xi = 0.34, V_0 = 0.07, \rho^* = 0 $ & Detlefsen \& H\"{a}rdle~\cite{Detlefsen2007} &          0 &     0.2731 &    -0.1773 \\

        IV &     Heston & $ \kappa =0.9, \delta = 0.04, \xi = 0.3, V_0 = 0.04, \rho^* = 0 $ & Andersen~\cite{Andersen2007b} &          0 &     0.2731 &    -0.1773 \\

         V &       CGMY & $ C = 1, G = 5, M = 10, Y = 0.5 $ & Madan \& Yor~\cite{Madan2008} &    -2.0000 &     0.7264 &    -0.1320 \\
\hline
\hline
\end{tabular}
\smallskip
\caption{\tiny Values of the market-dependent parameter ($\mu_{TC} $), and the maximum and minimum adjusted log-moneyness ($x_{TC}^{max} $ and $x_{TC}^{min} $) for five sets of parameters found in the literature (* we have set $ \rho $ equal to zero, as our framework only considers the zero-correlation Heston model, even though it was not zero in the original papers). \label{Tab:ChosenParameters}}
\end{center}
\end{table}
\normalsize \\
One can see from the table that the values for $\mu_{TC} $, $x_{TC}^{max} $ and
$x_{TC}^{min} $ vary substantially between different models and parameters.
Note that $ \mu_{TC} $ and $ \omega $ are both zero in the case
of the zero-correlation Heston model, implying that $\mu_{TC} $ and $x_{TC} $
do not depend on the model parameters at all. From here on we refer to these sets of
parameters as Case I-V, respectively.

\subsection{Cumulative distribution function of $\sigma^2 Z_T $}
\label{Sec:pdfofM}
\begin{figure}[t!]
\centering
\subfigure[Variance Gamma model]
{\label{Fig:CDFgamma}
    \includegraphics[height=6cm, width=8cm]{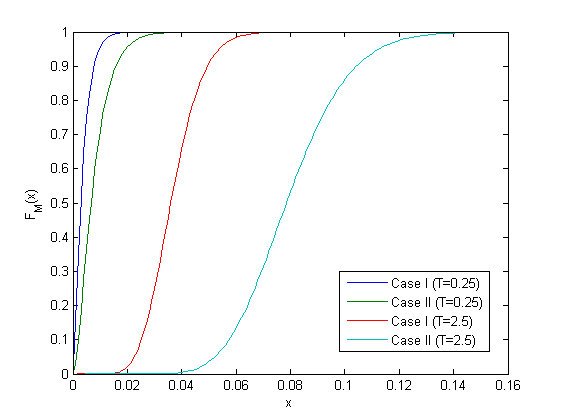}} 
\subfigure[Heston model]
{\label{Fig:CDFheston}
    \includegraphics[height=6cm, width=8cm]{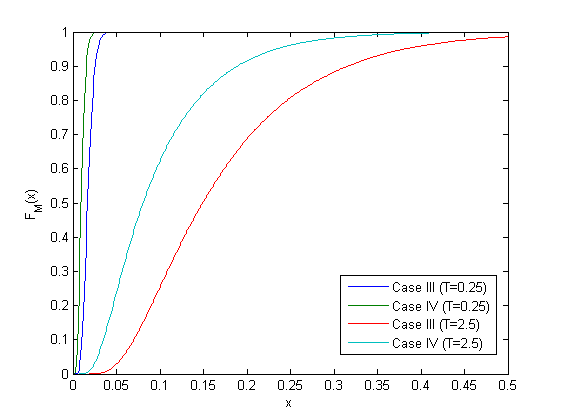}} 
    \caption{Plotted are the cumulative distribution functions of the scaled time-change $\sigma^2 Z_T $ for maturities $ T = 0.25 $ and $ T=2.5 $ for (a) the VG model with parameters as in Case I and II, and (b) the Heston model with parameters as in Case III and IV.\label{Fig:CDF}}
\end{figure}

\noindent
To choose appropriate values for $ a $ and $ b$, the truncation values of $ v$, we analyse the cdf of $ Z_T $ for Cases I - IV of Table \ref{Tab:ChosenParameters}. As stated in Section~\ref{Sec:VG}, $ Z_T $ is gamma-distributed ($ Z_T \sim G(\frac{T}{\nu} , \nu)$) for Cases I and II. By the scaling property of the gamma distribution, we obtain that $\sigma^2 Z_T \sim G(\frac{T}{\nu} , \sigma^2 \nu) $, with probability density function
\begin{eqnarray}
f_{\sigma^2 Z_T} \left(x, \frac{T}{\nu} ,  \sigma^2 \nu \right) = \frac{x^{\frac{t}{\nu}-1} e^{- \frac{x \sigma^2}{\nu}}}{(\sigma^2 \nu)^{\frac{t}{\nu}} \Gamma(\frac{t}{\nu})}.
\label{Eqn:pdfVGMt}
\end{eqnarray}
Plotting the cdf for Cases I and II with maturity $ T = 0.25 $ and $ T=2.5 $ results in Figure \ref{Fig:CDFgamma}.
For the CGMY model and the Heston model, we need to simulate $ Z_T $
to estimate its cdf as there are no closed-form expressions.
Madan \& Yor~\cite{Madan2008} describe a method to
simulate $ Z_T $ in the case of the CGMY model.
Glasserman \& Kim~\cite{Glasserman2009} describe a method to simulate $ Z_T $ in the case of the Heston model.
In particular, given $ V_0 $ in the Heston model, one can simulate $ V_T $, the endpoint of the variance process, exactly.
This follows since the transition law of the square-root diffusion (CIR diffusion) is a scaled noncentral chi-square distribution with
\begin{eqnarray}
V_T = \frac{\xi^2 (1-e^{-\kappa T})}{4 \kappa } \chi'^2_{\gamma} \left( \frac{4 \kappa e^{-\kappa T}}{\xi^2 (1-e^{-\kappa T})} V_0 \right), \text{    } T > 0, \text{    }  \delta = \frac{4 \kappa \delta}{\xi^2},
\end{eqnarray}
where $ \chi'^2_{\gamma} ( \lambda ) $ denotes a noncentral chi-square random variable
with $ \gamma $ degrees of freedom and noncentrality parameter $ \lambda $.
Then, conditional on $ V_0 $ and $ V_T $, the distribution of $ Z_T = \int_0^T V_s ds $ is given by
Theorem $ 2.1 $ in Glasserman \& Kim~\cite{Glasserman2009} and can therefore be simulated easily.
We implemented this procedure and estimated the cdf numerically for Cases III and
IV. Results are represented in Figure \ref{Fig:CDFheston}. As can be seen,
the ranges of the cdfs vary widely between the different models and different
maturities (note that the scales of Figures \ref{Fig:CDFgamma} and \ref{Fig:CDFheston} are not the same).

To ensure that only $0.1\%$ of the mass of $ \sigma^2 Z_T $ lies in the regions $ [0,a] $ and $ [b,\infty] $, we set $a$ and $b$ equal to the $0.1\%$
and $99.9\%$ quantiles of $\sigma^2 Z_T$, that is, we solve for $x$
setting $ p $ equal to $ 0.001 $ and $ 0.999 $ in the following equation:
\begin{eqnarray}
F^{-1}_{\sigma^2 Z_T} (p) = \text{inf}\{ x : F_{\sigma^2 Z_T} (x) \geq p \}.
\label{Eqn:inversecdfVG}
\end{eqnarray}
For the gamma distribution, for example, this inverse is then given by
\begin{eqnarray}
F^{-1}_{\sigma^2 Z_T} \left( p, \frac{T}{\nu} ,  \sigma^2 \nu \right) =  \text{inf} \left\{ x : F_{\sigma^2 Z_T} \left(x, \frac{T}{\nu} ,  \sigma^2 \nu \right) \geq p \right\},
\label{Eqn:cdfVG}
\end{eqnarray}
where
\begin{eqnarray}
F_{\sigma^2 Z_T} \left(x, \frac{T}{\nu} ,  \sigma^2 \nu \right) = \int_0^x f_{\sigma^2 Z_T}(s) ds = \gamma \left( \frac{T}{\nu}, \frac{x}{\sigma^2 \nu} \right),
\label{Eqn:cdfVG}
\end{eqnarray}
and $ \gamma $ is the upper incomplete gamma function defined as
\begin{eqnarray}
\gamma(a,b) = \Gamma(a)^{-1} \int_0^b t^{a-1} e^{-t} dt.
\end{eqnarray}
For the Heston model, we use simulated values of $ F_{\sigma^2 Z_T}(x) $ to calculate $ a $ and $ b $ (and the same could be done for the CGMY model). Results for Cases I - IV are presented in Table \ref{tab:abExamples}, where we have again chosen maturities $ T = 0.25 $ and $ T = 2.5 $ as an example and have calculated $ a $ and $ b $ for each case separately. These numbers confirm that the region of interest will depend heavily on the model and maturity under consideration.
\tiny
\begin{table}[htbp]
\begin{center}
\begin{tabular}{lrrrr}
\hline
      Case & $ a\, (T=0.25) $ & $ b\, (T=0.25) $ & $ a\, (T=2.5) $ & $ b\, (T=2.5) $ \\
\hline
         I & $ 2.84 \times 10^{-5} $ &     0.0201 &     0.0141 &     0.0735 \\

        II & $ 1.48 \times 10^{-4} $ &     0.0382 &     0.0347 &     0.1491 \\

       III &     0.0048 &     0.0413 &     0.0286 &     0.7283 \\

        IV &     0.0021 &     0.2600 &     0.0130 &     0.4841 \\
\hline
\hline
\end{tabular}
\smallskip
\caption{\tiny Values of $ a $ and $ b $, the $0.1\%$ and $99.9\%$ quantiles of the cumulative distribution function of $\sigma^2 Z_T$, for Case I - IV with maturity $ T = 0.25 $ and $ T = 0.5 $. \label{tab:abExamples}}
\end{center}
\end{table}
\normalsize \\
Clearly, the more that is known about the possible range of $\sigma^2 Z_T$ and the smaller we are able to make the range $ [a,b] $, the easier it will be to find an appropriate rational approximation for this range. There
are also extreme parameter combinations for which the range of $ [a,b] $ is rather large and the method does not perform as well (that is if the time change $ (Z_t) $ is too wide). Also, for maturities smaller than $ T = 0.25 $ we found that the resulting $ a $ might
be too small for some of the
parameter combinations and models, and the rational
approximation may result in larger maximum errors.

\subsection{Error from the rational function approximation}
\label{Sec:ErrorOfRationalFunctionApproximation}
\noindent
The error of most approximations looks similar to the error displayed in Figure \ref{fig:MaxErrorExample1},
where we now choose $ a $ and $ b $ as described in the previous section.
Note that the support of the time change in Equation (\ref{Eqn:CTC}) is over the whole positive real axis.
The approximation in Equation (\ref{Eq:CTC2}) therefore leads to errors over $ [0,\infty) $ as well. We subdivide this region into $ [0,a] $, $ [a,b] $ and $ [b,\infty) $ and approximate the error for each region and for each of the five cases of Table \ref{Tab:ChosenParameters} separately.
The errors can be calculated as follows:
\begin{eqnarray*}
\epsilon_{[0,a]} = \int_0^a \left[ c_{BS} \left(z;x,\mu \right) - \left( A_0^{x,\mu} + \sum_{i=0}^n b_i e^{- i z} + \sum_{k=1}^L \left( w_k \sum_{j=1}^m A_j^{x,\mu} e^{B_j^{x,\mu} x_k} \right) e^{- x_k z} dy \right) \right] f_{\sigma^2 Z_t} (z) dz, \\
\epsilon_{[a,b]} = \int_a^b \left[ c_{BS} \left(z;x,\mu \right) - \left( A_0^{x,\mu} + \sum_{i=0}^n b_i e^{- i z} + \sum_{k=1}^L \left( w_k \sum_{j=1}^m A_j^{x,\mu} e^{B_j^{x,\mu} x_k} \right) e^{- x_k z} dy \right) \right] f_{\sigma^2 Z_t} (z) dz, \\
\epsilon_{[b,\infty]} = \int_b^{\infty} \left[ c_{BS} \left(z;x,\mu \right) - \left( A_0^{x,\mu} + \sum_{i=0}^n b_i e^{- i z} + \sum_{k=1}^L \left( w_k \sum_{j=1}^m A_j^{x,\mu} e^{B_j^{x,\mu} x_k} \right) e^{- x_k z} dy \right) \right] f_{\sigma^2 Z_t} (z) dz.	
\end{eqnarray*}
Calculating these errors for each of the $ 246 $ options of Cases I - IV separately results in Table \ref{tab:epsilons}, where we represent the maximum absolute error for each case.
\tiny
\begin{table}[t]
\begin{center}
\begin{tabular}{lrrr}
\hline
      Case & $ \epsilon_{[0,a]}^{max} $ & $ \epsilon_{[a,b]}^{max} $ & $ \epsilon_{[b,\infty]}^{max} $ \\
\hline
         I & $ 1.44 \times 10^{-6} $ & $ 1.13 \times 10^{-5} $ & $ 2.84 \times 10^{-6} $ \\

        II & $ 1.29 \times 10^{-6} $ & $ 2.70 \times 10^{-5} $ & $ 6.36 \times 10^{-6} $ \\

       III & $ 2.73 \times 10^{-10} $ & $ 5.78 \times 10^{-9} $ & $ 2.67 \times 10^{-10} $ \\

        IV & $ 5.93 \times 10^{-8} $ & $ 5.84 \times 10^{-5} $ & $ 2.88 \times 10^{-8} $  \\

\hline
\hline
\end{tabular}
\smallskip
\caption{\tiny Values of the maximum errors resulting from the rational function approximation for $246$ parameter combinations over the region $ [0,a] $, $ [a,b] $ and $ [b,\infty) $, which are denoted by $ \epsilon_{[0,a]}^{max} $, $ \epsilon_{[a,b]}^{max} $ and $ \epsilon_{[b,\infty]}^{max} $ respectively. \label{tab:epsilons}}
\end{center}
\end{table}
\normalsize

As can be seen, maximum logarithmic errors are of order $ -5 $ to $ -6 $. Even though these errors are already relatively small, we observed that most of the $ 246 $ errors are much smaller than these maximum errors for each case.
To illustrate this point,
we recalculate the errors shown in Table \ref{tab:epsilons}, deleting
the largest $ 25 $ values from each maximum absolute error
calculation. The results are given in Table \ref{tab:epsilons2},
where we denoted these errors by $ \epsilon^{max25} $.
\tiny
\begin{table}[t]
\begin{center}
\begin{tabular}{lrrr}
\hline
      Case & $ \epsilon_{[0,a]}^{max25} $ & $ \epsilon_{[a,b]}^{max25} $ & $ \epsilon_{[b,\infty]}^{max25} $ \\
\hline
         I & $ 2.62 \times 10^{-7} $ & $ 3.70 \times 10^{-6} $ & $ 8.23 \times 10^{-7} $ \\

        II & $ 9.32 \times 10^{-7} $ & $ 8.97 \times 10^{-6} $ & $ 3.39 \times 10^{-6} $ \\

       III & $ 4.96 \times 10^{-11} $ & $ 9.56 \times 10^{-10} $ & $ 1.08 \times 10^{-11} $ \\

        IV & $ 6.91 \times 10^{-11} $ & $ 3.67 \times 10^{-10} $ & $ 2.87 \times 10^{-11} $  \\

\hline
\hline
\end{tabular}
\smallskip
\caption{\tiny Values of the maximum errors resulting from the rational function approximation for $246$ parameter combinations over the region $ [0,a] $, $ [a,b] $ and $ [b,\infty) $ after deleting the $25$ largest errors for each calculation, which are denoted by $ \epsilon_{[0,a]}^{max25} $, $ \epsilon_{[a,b]}^{max25} $ and $ \epsilon_{[b,\infty]}^{max25} $ respectively. \label{tab:epsilons2}}
\end{center}
\end{table}
\normalsize
When pricing options, one could therefore delete $x$ values that result in errors that are too large and interpolate between the $x$ values prices that had smaller errors (cf. discussion in Section \ref{Sec:splining}).
From Table \ref{tab:epsilons2} we see that, especially for Cases III and IV (the Heston model), the deletion of values with large errors and consequent interpolating should yield a great improvement. We will detail
this procedure further in the next section and give error estimates in
Section \ref{Sec:NumericalResults}.

\section{Implied Volatility}
\label{Sec:ImpliedVolatility}
\noindent
The Black-Scholes formula is often used to express option quotes in terms
of the {\it implied volatility}: Given the option price and all parameters except $ \sigma $ in the Black-Scholes formula (\ref{Eqn:BSEquation}), one searches for the value of $ \sigma $ that solves the equation. This value of $ \sigma $ is referred to as implied volatility and we denote it by $ \sigma_{IV} $. Since there is no closed-form formula for $ \sigma_{IV} $, it is usually solved for by using a solver method, which is typically rather slow. To speed up the calculation Manaster \& Kohler~\cite{Manaster1982} developed a technique that provides a starting value and guarantees convergence. More recent methods, such as the methods developed by Brenner \& Subrahmanyam~\cite{Brenner1988}, Corrado \& Miller~\cite{Corrado1996} and especially J\"{a}ckel~\cite{Jaeckel2006}, are faster and more accurate. Adopting the method described in the current paper, we construct an approximation of the inverse function by a rational function, which results in a fast and accurate alternative to the standard solver methods.

We restate the normalisation of the Black-Scholes formula (\ref{Eqn:cBS}) with a slightly modified parameterisation:
\begin{eqnarray}
c_{IV} (v,x) = c_{BS} (v^2,x,0) =  N \left( \frac{x}{v} + \frac{v}{2} \right) - e^{-x} N \left( \frac{x}{v} - \frac{v}{2} \right). \label{Eqn:LisNormalizedCall}
\end{eqnarray}
Any call option price can then be written in terms of
this normalised formula as follows:
\begin{eqnarray}
C_{BS} (S_0,K,T,\sigma,r,q) &=& S_0 e^{-q T} c_{IV} \left(\sigma \sqrt{T}, \text{log} \left( \frac{S_0 e^{(r-q)T}}{K} \right) \right).
\label{Eqn:LisRewrittenCall}
\end{eqnarray}

Since the normalised Black-Scholes formula $c_{IV}$ is a function of only two variables, finding a suitable rational approximation
for any parameter combination becomes easier. Note that the question of finding $ \sigma_{IV} $ reduces to finding $ v_{IV} $  in Equation~(\ref{Eqn:LisNormalizedCall}), where it follows from (\ref{Eqn:LisRewrittenCall}) that
\begin{eqnarray*}
c_{IV} &=& \frac{C_{BS} (S_0,K,T,\sigma,r,q)}{S_0 e^{-q T}}, \\
x &=& \text{log} \left( \frac{S_0 e^{(r-q)T}}{K} \right),
\end{eqnarray*}
which are both given constants when searching for the implied volatility. We therefore need to solve for the value of $ v $ that solves Equation (\ref{Eqn:LisNormalizedCall}) for known values $ c_{IV} $ and $ x $. Once we obtain $ v_{IV} $, $ \sigma_{IV} $ is determined by $ \sigma_{IV}  =  \frac{v_{IV}}{\sqrt{T}} $. It should also be noted that one can restrict, without loss of generality, to the case of call options,
as the case of put options follows on account of the put-call parity as in Section \ref{Sec:VanillaCallOption}.
In addition, reasoning as in the proof of Corollary
\ref{Cor:SymmetrycBS}, the following relation can be seen to hold true:
\begin{eqnarray}
c_{IV} (v,-x) = e^x c_{IV} (v,x) + 1 - e^x,\qquad x\in\mbb R, v\in\mathbb R_+,
\end{eqnarray}
so that the following relation holds for the implied volatility:
\begin{eqnarray}
v_{IV} (c_{IV},x) = v_{IV} (e^x c_{IV} + 1 - e^x,-x).
\end{eqnarray}
As a consequence, one can concentrate on the negative half-line $ x \in \mathbb{R}^- $.
All results presented in this section so far can be found in a
recent article by Li~\cite{Li2008}, who concentrates
on approximating the two-dimensional Black-Scholes implied volatility formula $ v_{IV} $  by a
single two-dimensional rational approximation. Li approximates the inverse of Equation (\ref{Eqn:LisNormalizedCall}) for $ -0.5 \leq x \leq 0$ and $ c_{LB}(x) \leq c_{IV} \leq c_{UB}(x) $ where
\begin{eqnarray*}
c_{LB}(x) &=& \frac{-0.00424532412773x + 0.00099075112125x^2}{1 + 0.26674393279214x + 0.03360553011959x^2}, \\
c_{UB}(x) &=& \frac{0.38292495908775 + 0.31382372544666x+0.07116503261172x^2}{1 + 0.01380361926221x + 0.11791124749938x^2}.
\end{eqnarray*}
This means that bounds are widest for $ x = 0 $ and become narrower as $ x $ decreases to $ -0.5 $.
These bounds are well chosen by Li, as it becomes more and more difficult to approximate the function well for small values of $ x $.
Still, it should be noted that the range is rather wide.

Note that the inverse is a function of two variables ($c_{IV}$,$x$). Li therefore uses a two-parameter rational approximation of the form
\begin{eqnarray}
v_{IV} (c_{IV},x) \approx v_{RA}^{Li} (c_{IV},x) = p_1 x + p_2 \sqrt{c_{IV}} + p_3 c_{IV} + \frac{\sum_{1 \leq i+j \leq 4} n_{i,j} x^i \sqrt{c_{IV}}^j}{1+ \sum_{1 \leq i+j \leq 4} m_{i,j} x^i \sqrt{c_{IV}}^j},
\end{eqnarray}
and obtains the approximation parameters using the downhill
simplex search method. Without further adjustments (like a Newton-Raphson polishing), Li states a maximum error of $ 3.3 \times 10^{-3} $ for $ v_{IV} $ over this region. It should be noted that the error is of similar magnitude for many parameter combinations and that one still needs to divide by $ \sqrt{T} $ to obtain $ \sigma_{IV} $ (making the error even bigger for short maturities). It should also be stressed that the approximation of Li applies to all combinations of $ c_{IV} $ and $ x $ (within the bounds) and a total of $ 31 $ parameters is sufficient for any calculation. Having worked with the downhill simplex search method ourselves, we note that this is a remarkable result and that it should be used wherever this error bound is acceptable.

\begin{figure}[t!]
\centering
\scalebox{0.5}
{\includegraphics{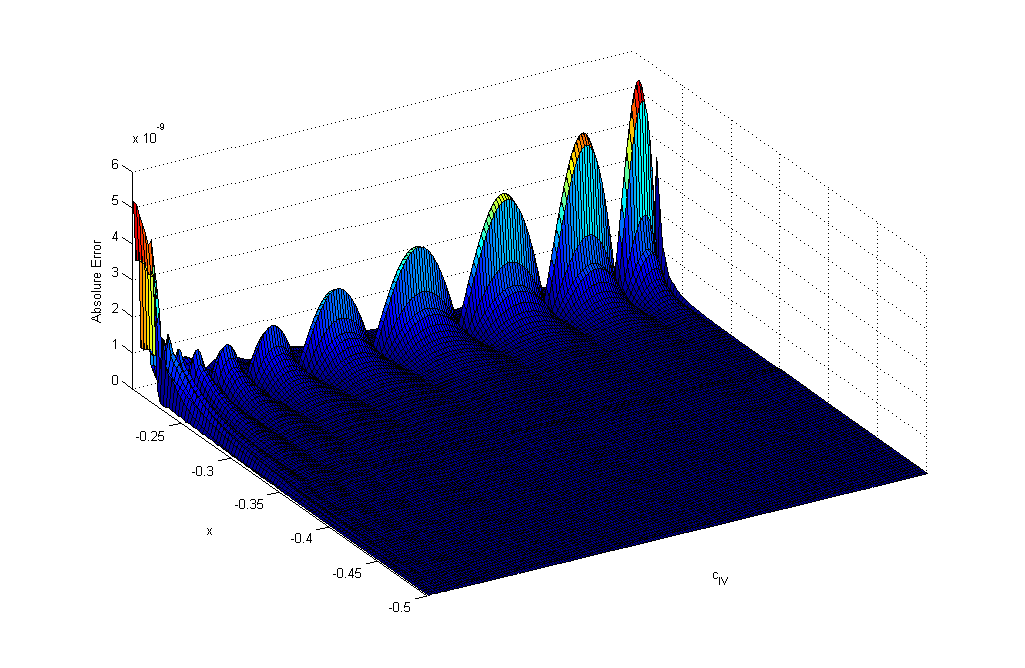}}
\caption{Displayed is the error in the normalised Black-Scholes implied volatility, $ v_{imp} $, for $ x \in [-0.5;-0.2] $ and $ c_{LB}(x) \leq c_{IV} \leq c_{UB}(x) $ resulting from the rational function approximations. \label{fig:ErrorRAvol}}
\end{figure}

Rather than the ambitious attempt to approximate the whole function with one set of parameters, we repeat the approximation for $ 105 $ values of $ x $ given by $ x = 0, - 0.0025, -0.005, -0.0075 , ... , -0.02 $ and $ x = -0.025, -0.03, -0.035, ... , -0.5 $. For each of these $ x $ values we use a rational function approximation of the following form:
\begin{eqnarray}
v_{IV} (c_{IV},x) \approx v_{RA}^{p} (c_{IV},x) = \frac{\sum_{i = 0}^p n_i^x \sqrt{c_{IV}}^i}{\sum_{i =0}^p m_i^x \sqrt{c_{IV}}^i},
\end{eqnarray}
where $ p $ varies between $ 7 $ and $ 9 $ for different values of $ x $ \footnote{For some values of $x$, the approximation is harder and we choose degree $ 9 $ for those. For other values, degree $ 7 $ is sufficient to obtain the desired error bound. Note that rational approximations of degree $9$ can still be performed easily with double precision. One can improve all the approximations performed in this paper by switching to higher order rational approximations, although, high precision arithmetic would become necessary for degrees of $ 13 $ and higher.}. The approximation parameters $ n_i^x $ and $ m_i^x $, where these all depend on $ x $, can be obtained again using the rational Chebyshev approximation (discussed in Appendix \ref{Sec:App:RationalApproximations}). However, we found that here it is advantageous to use the iterative algorithm discussed in Section 5.13 of Press \textit{et al.}~\cite{Press2002}, using a maximum of six iterations to get closer to the so-called ideal ``minimax" solution. Once the approximation has been performed for each value of $ x $, approximation parameters are stored to the computer and do not need to be recalculated, as detailed in Section \ref{Sec:splining}. Note that these approximations neither depend on the model nor on the market parameters and can therefore be used for any option.
To calculate $ v_{IV} $ for any value $ -0.5 \leq x \leq 0$, one only needs to interpolate between the previously calculated values (using linear interpolation or a cubic spline, for example).
We observe that the maximum absolute error for $ -0.5 \leq x \leq -0.0075 $ is $ 8.55 \times 10^{-7} $, whereas the maximum absolute error for $ -0.0075 < x \leq 0 $ is $ 5.54 \times 10^{-5} $. For most of the input values, the absolute error is even smaller than $ 1 \times 10^{-8} $ and we plot the error for $ x \in [-0.5;-0.2] $ in Figure \ref{fig:ErrorRAvol}. It is clear that the algorithm used is able to spread the error almost evenly for any of the $ x $ values and that all errors are rather small.

We would like to stress that this method, like the method presented by Li, does not require any approximation when calculating the implied volatility online. All approximations are done once and are stored to the hard-drive. The parameters one needs to store are all real and the calculation could easily be performed in Microsoft Excel, for example, without further adjustments.
Evaluating the implied volatility for any model and any set of parameters (within the bounds) can then be carried out extremely quickly via interpolation. The region that is well approximated by this method is wide and applies to most parameter combinations and options. The region where this method struggles is the region where most methods do: options very close to maturity with only a few hours or days remaining. In these cases, asymptotic methods are
appropriate (see e.g. Gao \& Lee~\cite{Gao2011}).
If some of the options being priced do not fall within the defined bounds (the bound on $ x $ should hardly ever be an issue, but the bounds on $ c_{IV} $ might be), there is no harm in using your favourite solver method for these few cases. For all other cases, the method presented here should yield a great improvement of computational efficiency and is easy to implement.

\section{Numerical results: Case Studies}
\label{Sec:NumericalResults}
\noindent
This section comprises the numerical results of the rational approximation method. Section \ref{Sec:NumericalResultsOptionPrices} compares numerical results for option prices of the developed rational approximation method to those obtained by employing the FFT method. Section \ref{Sec:ComparisonOfSpeed} compares the speeds of both methods and Section \ref{Sec:NumericalResultsImpliedVolatility} contains a comparison of accuracy and speed for the implied volatility calculations.

\subsection{Comparing Option Prices to the FFT}
\label{Sec:NumericalResultsOptionPrices}
\noindent
In the literature, other methods have been developed to
evaluate options numerically when the asset dynamics are assumed to be
different than in the Black-Scholes setting. Monte Carlo methods and finite
difference schemes are two widely used examples. In the case of asset dynamics
following Equations (\ref{Eqn:StockPriceProcess}) - (\ref{Eqn:TimeChangedBM}), the
standard methodology to evaluate vanilla options is to use Fourier-based option pricing methods. The Fourier transform can be efficiently computed using the standard FFT algorithm, and one numerical Fourier transform gives option prices for a whole range of strikes with the same maturity. This method was introduced to finance by Carr \& Madan~\cite{Carr19982} and later extended and generalised by a number of authors (e.g. Eberlein \textit{et al.}~\cite{Eberlein2010}, Lee~\cite{Lee20042}, Lewis~\cite{Lewis2001} and Raible~\cite{Raible2000}). More recently, other efficient versions of the Fourier method have been developed as well. However, as the standard FFT method is well-known and widely used in practice, we compare the rational approximation approach presented here to the standard FFT method, employing the Carr-Madan algorithm with the recommended values of parameters
(dampening parameter $\alpha=1.5$, step-size $\eta=0.25$ and the number of steps $N=2048$). Note that for the FFT method, one usually requires the characteristic function of the log of the stock price at time $t$ (where the stock price at time $t$ is assumed to be as in Equation~(\ref{Eqn:StockPriceProcess})), which is
\begin{eqnarray}
\mathbb{E}[ e^{i z \text{log} (S_t)}] = e^{i z ( \text{log} (S_0) + (r-q-\phi(-i)) t)}  \mathbb{E}^{\mbb Q}[ e^{i z X_t}].
\label{Eqn:CharFuncLogSt}
\end{eqnarray}
As can be seen, one therefore requires the Fourier transform of $ X_t$ as defined in Equation (\ref{Eqn:characteristicfunction}) rather than the Laplace transform of $ Z_t $. The general formula to obtain the resulting characteristic function $ \Phi_{X_t}(z) $ for $ X_t $  from that of $ Z_t $ is
\begin{eqnarray}
\mathbb{E}^{\mbb Q} [e^{i z X_t}] &=& \mathbb{E}^{\mbb Q} [ e^{i z (\sigma W_{Z_t} + \theta Z_t ) } ]  \label{Eqn:CharacteristicFunctionOfX} \\
 &=& \mathbb{E}^{\mbb Q} [ e^{ ( i z \theta - \frac{z^2 \sigma^2}{2}) Z_t } ], \ \ z \in \mathbb{R} \nonumber,
\end{eqnarray}
for all three models discussed above. From this it follows that, for example, the characteristic function of the log-price
$X_t$ in the VG model is equal to
\begin{eqnarray}
\mathbb{E}^{\mbb Q} [e^{i z X_t}] = (1 - \nu z i \theta + \sigma^2 \nu z^2 / 2 )^{- \frac{t}{\nu}}.
\label{Eqn:CharFunVG}
\end{eqnarray}

We next compare prices of vanilla call options computed from both methods for the three models discussed in Section \ref{Sec:LevyProcessInMathematicalFinance}. For each of the models, we select one set of market parameters from Table \ref{Tab:ChosenParameters} and compare plain vanilla call option prices for strikes $ K \in [0.8, 1.2] $ and maturities $ T \in [0.25, 2.5]$. Tables \ref{tab:DifferencesVG} - \ref{tab:DifferencesHeston} give absolute differences between the option price obtained by the FFT method and the rational approximation method. In particular, Table \ref{tab:DifferencesVG} represents errors for Case I, Table \ref{tab:DifferencesCGMY} represents errors for Case V and Table \ref{tab:DifferencesHeston} represents errors for Case III. In these three tables, we set $ S_0 = 1 $, $ r = 0.03$ and $ q = 0.01$. As can be seen, the maximum values attained  in Table \ref{tab:DifferencesVG}, \ref{tab:DifferencesCGMY} and \ref{tab:DifferencesHeston} are $ 1.61 \times 10^{-5} $, $ 1.14 \times 10^{-6}$ and $8.59 \times 10^{-8}$, respectively. \\
For all three tables, we precalculated $ 30 $ values of $ x_{TC} $ equally spaced between $x_{TC}^{max} $ and $x_{TC}^{min} $ such that strikes $ K \in [0.8, 1.2] $ and maturities $ T \in [0.25, 2.5]$ are included. We iterate over ten different values for $a$ and $b$ and calculate errors for each of the ten cases, rather than calculating the value for $ a $ and $ b $ with the help of the cdf, as this procedure is much quicker. We take those values of $ a $ and $ b $ that lead to errors smaller than $ 1 \times 10^{-6}$ and delete $ x $ values for which no such combination can be found. In Equation (\ref{Eqn:CorrectionTerm}), $ n $ was set to $ 7 $. The rational Chebyshev approximation in Equation~(\ref{Eqn:RAc}) starts with degree $ 6 $ and increases to degree $ 8 $ to avoid positive roots, and the Gauss-Legendre quadrature in Equation (\ref{Eqn:GaussianQuad}) was performed with $ 500 $ points, where $ c = 0 $ and $ d = 7000 $.

{\bf Remark.} Note that also these degrees (that is rational approximations with degree 6 and 8) can be easily performed with double precision. If one is interested in improving on the error bounds given here, it might be worth investigating how switching to higher order rational approximations, which will require high precision arithmetics, influences the errors. This could be particularly relevant for the few approximations performed offline, for which computational effort is not the main concern. We also leave a detailed comparison of different Gaussian quadrature methods for further research.

\tiny
\begin{table}[!h!]
\begin{center}
\begin{tabular}{rrrrrrr}
\hline
         K &   T = 0.25 &    T = 0.5 &      T = 1 &    T = 1.5 &      T = 2 &    T = 2.5 \\
\hline
     0.80 &   3.89E-06 &   1.02E-07 &   9.25E-06 &   3.60E-06 &   6.82E-06 &   8.25E-06 \\

      0.81 &   3.46E-06 &   3.06E-08 &   9.54E-06 &   5.13E-06 &   4.26E-06 &   8.35E-06 \\

      0.82 &   1.04E-06 &   4.47E-08 &   5.27E-06 &   6.65E-06 &   1.99E-06 &   8.90E-06 \\

      0.83 &   6.84E-07 &   1.80E-07 &   3.78E-07 &   6.89E-06 &   1.11E-06 &   9.57E-06 \\

      0.84 &   6.39E-07 &   2.81E-07 &   1.28E-06 &   5.58E-06 &   1.88E-06 &   1.01E-05 \\

      0.85 &   3.98E-07 &   1.50E-07 &   5.89E-07 &   2.80E-06 &   3.17E-06 &   1.03E-05 \\

      0.86 &   1.65E-06 &   3.58E-07 &   2.46E-07 &   5.73E-07 &   3.75E-06 &   1.02E-05 \\

      0.87 &   2.17E-06 &   3.92E-07 &   3.03E-07 &   2.90E-06 &   3.23E-06 &   9.60E-06 \\

      0.88 &   7.26E-07 &   8.94E-08 &   2.93E-09 &   2.86E-06 &   2.16E-06 &   8.62E-06 \\

      0.89 &   1.13E-06 &   1.84E-07 &   5.28E-08 &   1.27E-06 &   9.45E-07 &   7.26E-06 \\

      0.90 &   4.48E-07 &   3.90E-07 &   4.83E-08 &   1.70E-07 &   3.16E-07 &   5.73E-06 \\

      0.91 &   3.54E-06 &   4.24E-07 &   1.03E-07 &   4.26E-07 &   1.88E-06 &   4.24E-06 \\

      0.92 &   4.12E-06 &   3.28E-07 &   9.98E-08 &   1.27E-07 &   4.12E-06 &   3.09E-06 \\

      0.93 &   1.65E-06 &   6.39E-08 &   7.38E-08 &   1.70E-07 &   6.21E-06 &   2.81E-06 \\

      0.94 &   4.24E-07 &   9.95E-08 &   1.89E-08 &   1.88E-07 &   6.89E-06 &   3.24E-06 \\

      0.95 &   3.07E-07 &   1.14E-07 &   5.36E-10 &   1.11E-08 &   5.42E-06 &   3.75E-06 \\

      0.96 &   7.42E-07 &   1.11E-07 &   7.96E-08 &   3.95E-08 &   2.58E-06 &   4.01E-06 \\

      0.97 &   4.44E-07 &   8.60E-08 &   1.25E-07 &   5.21E-08 &   2.61E-08 &   3.73E-06 \\

      0.98 &   1.54E-06 &   3.07E-07 &   6.27E-08 &   6.28E-08 &   7.55E-07 &   2.61E-06 \\

      0.99 &   2.87E-06 &   4.47E-07 &   1.73E-07 &   2.01E-07 &   4.55E-07 &   7.06E-07 \\

      1.00 &   2.06E-06 &   6.61E-08 &   3.40E-08 &   1.42E-07 &   4.76E-08 &   1.59E-06 \\

      1.01 &   2.35E-06 &   1.05E-06 &   1.01E-07 &   1.41E-07 &   3.10E-07 &   3.57E-06 \\

      1.02 &   4.92E-06 &   2.06E-06 &   1.12E-07 &   1.41E-07 &   1.77E-07 &   4.15E-06 \\

      1.03 &   8.92E-06 &   9.71E-07 &   1.16E-08 &   4.51E-10 &   2.02E-08 &   2.98E-06 \\

      1.04 &   1.61E-05 &   2.41E-07 &   1.77E-07 &   9.86E-08 &   1.85E-08 &   1.21E-06 \\

      1.05 &   4.25E-07 &   9.88E-07 &   2.58E-08 &   1.10E-07 &   5.21E-09 &   1.63E-07 \\

      1.06 &   1.98E-06 &   7.80E-07 &   1.16E-07 &   5.74E-08 &   2.82E-08 &   6.32E-07 \\

      1.07 &   3.00E-06 &   3.22E-08 &   1.09E-07 &   1.31E-07 &   1.40E-07 &   3.75E-07 \\

      1.08 &   2.18E-06 &   3.68E-07 &   1.19E-08 &   3.58E-08 &   1.12E-07 &   5.33E-08 \\

      1.09 &   2.63E-06 &   8.86E-07 &   1.28E-07 &   3.46E-08 &   9.61E-08 &   1.21E-07 \\

      1.10 &   1.95E-06 &   3.44E-07 &   1.33E-08 &   7.36E-08 &   1.11E-07 &   5.65E-08 \\

      1.11 &   2.70E-06 &   1.49E-06 &   4.10E-08 &   1.82E-09 &   2.33E-08 &   1.63E-09 \\

      1.12 &   2.05E-06 &   7.31E-07 &   5.48E-08 &   5.10E-08 &   1.95E-08 &   7.60E-08 \\

      1.13 &   9.87E-07 &   1.14E-07 &   3.67E-09 &   4.54E-08 &   6.25E-08 &   7.01E-08 \\

      1.14 &   5.68E-08 &   6.47E-07 &   3.83E-08 &   1.07E-10 &   3.08E-08 &   8.57E-10 \\

      1.15 &   1.01E-06 &   3.59E-07 &   7.05E-09 &   3.31E-08 &   4.20E-08 &   2.93E-08 \\

      1.16 &   1.27E-06 &   6.50E-08 &   2.64E-08 &   1.38E-08 &   4.78E-08 &   7.66E-08 \\

      1.17 &   1.03E-06 &   9.28E-08 &   8.06E-09 &   2.44E-09 &   4.45E-09 &   7.42E-08 \\

      1.18 &   9.50E-08 &   7.57E-08 &   3.20E-08 &   1.53E-08 &   5.20E-09 &   7.77E-08 \\

      1.19 &   1.10E-06 &   1.07E-07 &   6.20E-08 &   8.27E-09 &   1.35E-08 &   6.56E-08 \\

      1.20 &   2.17E-07 &   6.09E-09 &   3.24E-09 &   9.02E-09 &   1.73E-09 &   1.86E-08 \\
\hline
\hline
\end{tabular}
\smallskip
\caption{\tiny Absolute differences between call option prices in the VG model computed using FFT method and 
rational function method for varying strikes and maturities, with spot $S_0=1$, interest rate $ r = 0.03$, 
and dividend yield $ q = 0.01$. Parameters of the VG model were fixed as
$ \sigma = 0.1213 $, $ \nu = 0.1686 $ and $ \theta = -0.1436 $ (Case I of Table \ref{Tab:ChosenParameters}).  \label{tab:DifferencesVG}}
\end{center}
\end{table}

\begin{table}[!h!]
\begin{center}
\begin{tabular}{rrrrrrrr}
\hline
         K &   T = 0.25 &    T = 0.5 &      T = 1 &    T = 1.5 &      T = 2 &    T = 2.5 \\
\hline
    0.80 &   3.89E-06 &   1.02E-07 &   9.25E-06 &   3.60E-06 &   6.82E-06 &   8.25E-06 \\

      0.81 &   3.46E-06 &   3.06E-08 &   9.54E-06 &   5.13E-06 &   4.26E-06 &   8.35E-06 \\

      0.82 &   1.04E-06 &   4.47E-08 &   5.27E-06 &   6.65E-06 &   1.99E-06 &   8.90E-06 \\

      0.83 &   6.84E-07 &   1.80E-07 &   3.78E-07 &   6.89E-06 &   1.11E-06 &   9.57E-06 \\

      0.84 &   6.39E-07 &   2.81E-07 &   1.28E-06 &   5.58E-06 &   1.88E-06 &   1.01E-05 \\

      0.85 &   3.98E-07 &   1.50E-07 &   5.89E-07 &   2.80E-06 &   3.17E-06 &   1.03E-05 \\

      0.86 &   1.65E-06 &   3.58E-07 &   2.46E-07 &   5.73E-07 &   3.75E-06 &   1.02E-05 \\

      0.87 &   2.17E-06 &   3.92E-07 &   3.03E-07 &   2.90E-06 &   3.23E-06 &   9.60E-06 \\

      0.88 &   7.26E-07 &   8.94E-08 &   2.93E-09 &   2.86E-06 &   2.16E-06 &   8.62E-06 \\

      0.89 &   1.13E-06 &   1.84E-07 &   5.28E-08 &   1.27E-06 &   9.45E-07 &   7.26E-06 \\

      0.90 &   4.48E-07 &   3.90E-07 &   4.83E-08 &   1.70E-07 &   3.16E-07 &   5.73E-06 \\

      0.91 &   3.54E-06 &   4.24E-07 &   1.03E-07 &   4.26E-07 &   1.88E-06 &   4.24E-06 \\

      0.92 &   4.12E-06 &   3.28E-07 &   9.98E-08 &   1.27E-07 &   4.12E-06 &   3.09E-06 \\

      0.93 &   1.65E-06 &   6.39E-08 &   7.38E-08 &   1.70E-07 &   6.21E-06 &   2.81E-06 \\

      0.94 &   4.24E-07 &   9.95E-08 &   1.89E-08 &   1.88E-07 &   6.89E-06 &   3.24E-06 \\

      0.95 &   3.07E-07 &   1.14E-07 &   5.36E-10 &   1.11E-08 &   5.42E-06 &   3.75E-06 \\

      0.96 &   7.42E-07 &   1.11E-07 &   7.96E-08 &   3.95E-08 &   2.58E-06 &   4.01E-06 \\

      0.97 &   4.44E-07 &   8.60E-08 &   1.25E-07 &   5.21E-08 &   2.61E-08 &   3.73E-06 \\

      0.98 &   1.54E-06 &   3.07E-07 &   6.27E-08 &   6.28E-08 &   7.55E-07 &   2.61E-06 \\

      0.99 &   2.87E-06 &   4.47E-07 &   1.73E-07 &   2.01E-07 &   4.55E-07 &   7.06E-07 \\

      1.00 &   2.06E-06 &   6.61E-08 &   3.40E-08 &   1.42E-07 &   4.76E-08 &   1.59E-06 \\

      1.01 &   2.35E-06 &   1.05E-06 &   1.01E-07 &   1.41E-07 &   3.10E-07 &   3.57E-06 \\

      1.02 &   4.92E-06 &   2.06E-06 &   1.12E-07 &   1.41E-07 &   1.77E-07 &   4.15E-06 \\

      1.03 &   8.92E-06 &   9.71E-07 &   1.16E-08 &   4.51E-10 &   2.02E-08 &   2.98E-06 \\

      1.04 &   1.61E-05 &   2.41E-07 &   1.77E-07 &   9.86E-08 &   1.85E-08 &   1.21E-06 \\

      1.05 &   4.25E-07 &   9.88E-07 &   2.58E-08 &   1.10E-07 &   5.21E-09 &   1.63E-07 \\

      1.06 &   1.98E-06 &   7.80E-07 &   1.16E-07 &   5.74E-08 &   2.82E-08 &   6.32E-07 \\

      1.07 &   3.00E-06 &   3.22E-08 &   1.09E-07 &   1.31E-07 &   1.40E-07 &   3.75E-07 \\

      1.08 &   2.18E-06 &   3.68E-07 &   1.19E-08 &   3.58E-08 &   1.12E-07 &   5.33E-08 \\

      1.09 &   2.63E-06 &   8.86E-07 &   1.28E-07 &   3.46E-08 &   9.61E-08 &   1.21E-07 \\

      1.10 &   1.95E-06 &   3.44E-07 &   1.33E-08 &   7.36E-08 &   1.11E-07 &   5.65E-08 \\

      1.11 &   2.70E-06 &   1.49E-06 &   4.10E-08 &   1.82E-09 &   2.33E-08 &   1.63E-09 \\

      1.12 &   2.05E-06 &   7.31E-07 &   5.48E-08 &   5.10E-08 &   1.95E-08 &   7.60E-08 \\

      1.13 &   9.87E-07 &   1.14E-07 &   3.67E-09 &   4.54E-08 &   6.25E-08 &   7.01E-08 \\

      1.14 &   5.68E-08 &   6.47E-07 &   3.83E-08 &   1.07E-10 &   3.08E-08 &   8.57E-10 \\

      1.15 &   1.01E-06 &   3.59E-07 &   7.05E-09 &   3.31E-08 &   4.20E-08 &   2.93E-08 \\

      1.16 &   1.27E-06 &   6.50E-08 &   2.64E-08 &   1.38E-08 &   4.78E-08 &   7.66E-08 \\

      1.17 &   1.03E-06 &   9.28E-08 &   8.06E-09 &   2.44E-09 &   4.45E-09 &   7.42E-08 \\

      1.18 &   9.50E-08 &   7.57E-08 &   3.20E-08 &   1.53E-08 &   5.20E-09 &   7.77E-08 \\

      1.19 &   1.10E-06 &   1.07E-07 &   6.20E-08 &   8.27E-09 &   1.35E-08 &   6.56E-08 \\

      1.20 &   2.17E-07 &   6.09E-09 &   3.24E-09 &   9.02E-09 &   1.73E-09 &   1.86E-08 \\
\hline
\hline
\end{tabular}
\smallskip
\caption{\tiny Absolute differences between call option prices in the CGMY model computed using FFT method and
rational function method for varying strikes and maturities, with spot $S_0=1$, interest rate $ r = 0.03$,
and dividend yield $ q = 0.01$. Parameters of the CGMY model were fixed as $ C = 1 $, $ G = 5 $, $ M = 10 $ and $ Y = 0.5 $ 
(Case V of Table \ref{Tab:ChosenParameters}) \label{tab:DifferencesCGMY}}
\end{center}
\end{table}

\begin{table}[!h!]
\begin{center}
\begin{tabular}{rrrrrrrr}
\hline
         K &   T = 0.25 &    T = 0.5 &      T = 1 &    T = 1.5 &      T = 2 &    T = 2.5 \\
\hline
      0.80 &   3.33E-09 &   1.84E-10 &   6.13E-09 &   3.44E-09 &   1.86E-08 &   8.36E-10 \\

      0.81 &   8.19E-10 &   1.10E-09 &   1.13E-09 &   8.08E-09 &   6.10E-09 &   2.07E-08 \\

      0.82 &   1.36E-08 &   1.27E-09 &   2.09E-09 &   2.63E-09 &   5.13E-09 &   3.04E-10 \\

      0.83 &   5.67E-09 &   2.33E-09 &   5.48E-09 &   2.36E-09 &   2.79E-09 &   5.42E-10 \\

      0.84 &   1.56E-09 &   2.96E-09 &   3.44E-09 &   5.04E-09 &   2.83E-09 &   2.06E-09 \\

      0.85 &   9.52E-09 &   4.86E-09 &   3.01E-09 &   4.02E-09 &   3.81E-09 &   3.24E-09 \\

      0.86 &   2.05E-08 &   2.49E-09 &   6.01E-09 &   2.18E-09 &   3.86E-09 &   1.89E-09 \\

      0.87 &   6.16E-08 &   8.58E-09 &   2.88E-09 &   4.85E-09 &   1.28E-09 &   1.54E-09 \\

      0.88 &   6.76E-08 &   1.31E-08 &   2.91E-09 &   3.10E-09 &   1.31E-09 &   4.22E-09 \\

      0.89 &   7.48E-08 &   5.79E-09 &   5.69E-09 &   1.55E-09 &   1.39E-09 &   1.42E-08 \\

      0.90 &   7.27E-08 &   1.21E-08 &   1.56E-09 &   4.20E-09 &   5.16E-09 &   2.93E-08 \\

      0.91 &   7.25E-08 &   1.62E-08 &   3.57E-09 &   2.19E-09 &   4.17E-09 &   4.55E-08 \\

      0.92 &   6.47E-08 &   6.48E-09 &   7.01E-09 &   2.47E-09 &   3.39E-09 &   5.05E-08 \\

      0.93 &   3.50E-08 &   1.39E-08 &   2.83E-09 &   5.19E-09 &   4.23E-11 &   3.89E-08 \\

      0.94 &   3.58E-08 &   1.52E-08 &   5.63E-09 &   2.59E-09 &   3.64E-09 &   1.64E-08 \\

      0.95 &   2.68E-08 &   8.73E-10 &   6.79E-09 &   3.16E-09 &   1.23E-10 &   4.96E-09 \\

      0.96 &   6.57E-09 &   6.81E-09 &   6.62E-10 &   4.37E-09 &   3.68E-09 &   1.93E-08 \\

      0.97 &   4.79E-09 &   1.24E-08 &   3.55E-09 &   1.61E-09 &   5.66E-10 &   4.03E-08 \\

      0.98 &   3.66E-08 &   9.89E-09 &   7.21E-09 &   3.51E-09 &   2.20E-09 &   2.42E-08 \\

      0.99 &   5.90E-08 &   1.92E-08 &   6.14E-09 &   4.72E-09 &   3.67E-09 &   2.24E-09 \\

      1.00 &   6.93E-08 &   1.54E-08 &   7.81E-09 &   2.70E-09 &   3.50E-09 &   4.25E-09 \\

      1.01 &   6.18E-08 &   3.47E-09 &   4.50E-09 &   3.73E-09 &   1.70E-09 &   3.40E-10 \\

      1.02 &   5.41E-08 &   1.20E-08 &   4.80E-10 &   2.68E-09 &   2.01E-09 &   1.09E-09 \\

      1.03 &   3.89E-08 &   1.01E-08 &   4.10E-09 &   1.05E-09 &   1.18E-09 &   5.24E-09 \\

      1.04 &   1.10E-08 &   2.49E-09 &   4.23E-09 &   3.23E-09 &   4.80E-09 &   2.49E-08 \\

      1.05 &   3.85E-10 &   8.45E-09 &   2.75E-09 &   3.30E-09 &   1.55E-09 &   4.42E-08 \\

      1.06 &   6.61E-09 &   3.29E-09 &   4.35E-09 &   2.79E-09 &   2.01E-09 &   2.93E-08 \\

      1.07 &   5.51E-09 &   2.04E-09 &   2.28E-09 &   3.55E-09 &   4.32E-09 &   1.11E-08 \\

      1.08 &   1.41E-08 &   4.55E-09 &   1.28E-09 &   2.57E-09 &   4.39E-09 &   1.02E-08 \\

      1.09 &   2.97E-08 &   5.54E-09 &   3.69E-09 &   2.75E-09 &   2.16E-09 &   4.30E-09 \\

      1.10 &   4.24E-08 &   4.46E-09 &   2.55E-09 &   3.15E-09 &   1.91E-09 &   9.78E-10 \\

      1.11 &   6.83E-08 &   7.55E-09 &   1.94E-09 &   2.14E-09 &   3.31E-09 &   1.54E-09 \\

      1.12 &   8.59E-08 &   9.40E-09 &   2.83E-09 &   2.16E-09 &   1.79E-09 &   3.90E-09 \\

      1.13 &   8.15E-08 &   1.06E-08 &   2.39E-09 &   2.28E-09 &   3.13E-09 &   1.05E-09 \\

      1.14 &   7.68E-08 &   8.75E-09 &   2.28E-09 &   1.76E-09 &   3.56E-09 &   3.15E-08 \\

      1.15 &   7.10E-08 &   7.58E-09 &   2.09E-09 &   1.85E-09 &   2.69E-10 &   4.10E-08 \\

      1.16 &   7.37E-08 &   8.91E-09 &   1.68E-09 &   1.69E-09 &   1.21E-09 &   1.68E-08 \\

      1.17 &   6.19E-08 &   9.94E-09 &   1.94E-09 &   1.62E-09 &   2.14E-09 &   7.82E-09 \\

      1.18 &   7.01E-09 &   1.25E-09 &   2.95E-09 &   1.63E-09 &   6.07E-09 &   2.41E-08 \\

      1.19 &   5.69E-08 &   2.98E-10 &   1.37E-09 &   1.81E-09 &   6.81E-09 &   5.02E-08 \\

      1.20 &   2.40E-09 &   3.74E-08 &   3.94E-10 &   1.78E-09 &   5.95E-09 &   5.77E-08 \\
\hline
\hline
\end{tabular}
\smallskip
\caption{ \tiny Absolute differences between call option prices in the zero-correlation Heston model 
computed using FFT method and
rational function method for varying strikes and maturities, with spot $S_0=1$, interest rate $ r = 0.03$,
and dividend yield $ q = 0.01$. Parameters in the Heston model were fixed as $ \kappa = 0.87 $, $\delta = 0.07$, $ \xi = 0.34$ ,$ V_0 = 0.07 $ and $ \rho = 0 $ (Case III of Table \ref{Tab:ChosenParameters}).  \label{tab:DifferencesHeston}}
\end{center}
\end{table}

\newpage
~
\newpage
~
\newpage

\normalsize 

\subsection{Comparison of speed}
\label{Sec:ComparisonOfSpeed}
The relative computational speeds of the FFT method (with $ 2048 $ points) and of the rational function method depend on the number
of strikes and the number of maturities that are considered.
In the FFT method, all strikes
can be computed in one go, given that they have the same maturity. The rational function approximation, on the other hand, approximates the normalised Black-Scholes
formula (\ref{Eqn:cBS}) for a given strike, which can then be used
for any maturity. Therefore, pricing relatively few strikes and
relatively many maturities
should be advantageous for the rational approximation method and vice versa.
Table \ref{tab:DifferencesTime} gives approximate computational
times for the two different methods,
for different numbers of strikes and maturities. The first row
resembles the example of Table \ref{tab:DifferencesVG}, where
we can see that the rational function approximation clearly
outperforms the FFT method. As can also be seen, the rational function
approximation is about five to ten times faster than the FFT method, depending
on the number of strikes and maturities that need to be priced.
The computational times given here were measured by Matlab (version R2009b)
on a Lenovo ThinkPad R60 (1.8 GHZ, 2GB RAM) and are given in seconds.
\tiny
\begin{table}[htbp]
\begin{center}
\begin{tabular}{rrrr}
\hline
\# Maturities &  \# Strikes & Time in sec. FFT & Time in sec. RA \\
\hline
        41 &          7 &      0.404 &      0.044 \\

         7 &         41 &      0.074 &      0.011 \\

       100 &        100 &      1.006 &      0.106 \\

       300 &        300 &      2.816 &      0.460 \\

       300 &          5 &      2.795 &      0.298 \\

         5 &        300 &      0.049 &      0.009 \\
\hline
\hline
\end{tabular}
\smallskip
\caption{\tiny Running times of the FFT method and rational function method, in seconds. \label{tab:DifferencesTime}}
\end{center}
\end{table}
\normalsize

\subsection{Numerical Results for Implied Volatility}
\label{Sec:NumericalResultsImpliedVolatility}
\noindent
We use our method presented in Section \ref{Sec:ImpliedVolatility} to determine the volatility surface of the CGMY model for $ T \in [0.1,1.0]$ (steps of $0.01$) and $ K \in [0.7,1.3] $ (steps of $0.01$) with $ C = 1 $, $ G = 5 $, $ M = 10 $ and $ Y = 0.5 $ (these parameters are taken from  Madan \& Yor~\cite{Madan2008}). Spot, interest rate and dividend yield are fixed at $ S_0 = 1$, $ r = 0.03 $ and $ q = 0.01 $. This gives a total of $ 5,551 $ options and results in a maximum error of $ 4.35 \times 10^{-7} $.
\begin{figure}[t]
\centering
\scalebox{0.5}
{\includegraphics{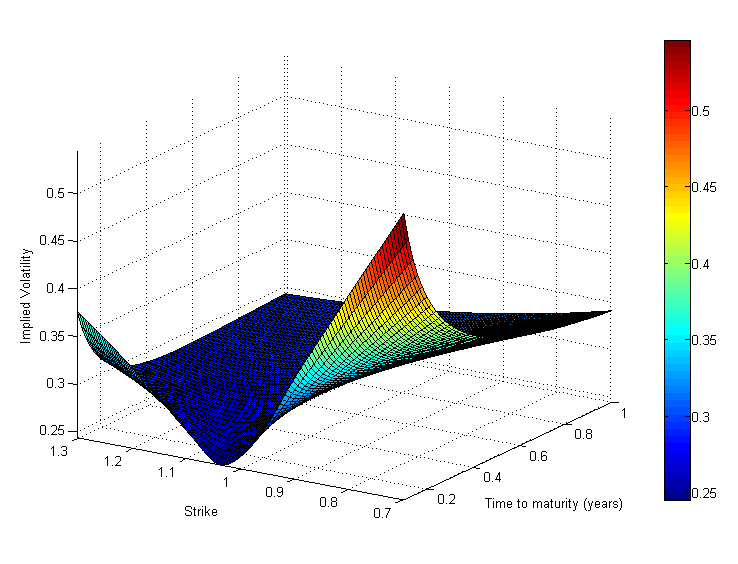}}
\caption{Plotted is the volatility surface corresponding to the prices of call options with maturities
$ T \in [0.1,1.0]$ and strikes $ K \in [0.7,1.3] $ under the CGMY model with $ C = 1 $, $ G = 5 $, $ M = 10 $ and $ Y = 0.5 $ (which are parameter values taken from Madan \& Yor~\cite{Madan2008}). Spot, interest rate and dividend yield are fixed at $ S_0 = 1$, $ r = 0.03 $ and $ q = 0.01 $.}
\end{figure}
As stated before, this procedure is faster than the usual solver methods. When comparing to Matlab's built-in solver method for implied volatility (setting accuracy to $ 1 \times 10^{-8}$), calculating these $ 5,551 $ implied volatilities takes $ 80.22 $ seconds with our rational function method and $ 320.42 $ seconds with Matlab's method.

\section{Conclusion}
\label{Sec:Conclusion}
\noindent
In this paper, we illustrate how to use rational function approximations to derive an alternative approach for calculating vanilla option prices and Black-Scholes implied volatilities. In particular, by approximating a normalisation of the Black-Scholes formula, a two-parameter function of one variable, we derive a pricing formula for any time-changed Brownian motion model. The method was found to be stable and fast, since it is possible to rewrite the rational approximation in terms of negative exponential moments (the Laplace transform) of the time change. The method can therefore be used for any model for which the Laplace transform of the time change is available in tractable form. Although rational Chebyshev approximations can be performed with little computational effort, we show that one can significantly speed up the method by pre-calculating the approximations and interpolating between them for the purpose of option pricing.

By way of illustration, we implemented the method for three popular models from the literature, and found that the method
compares favourably to the standard FFT with respect to speed and accuracy.

Additionally, we use the methodology to approximate the inverse of the Black-Scholes formula in order to compute implied volatilities,  and found this method to be significantly faster than existing methods. We are able to approximate the resulting two-dimensional function by a total of $ 105 $ rational approximations. Implied volatilities for a wide range of input parameters can then be obtained, for any given set of arbitrage-free vanilla option prices, by interpolating between the previously calculated approximations. The absolute error for this method is shown to be bounded by $ 5.5 \times 10^{-5} $, although this can be refined if necessary by choosing rational approximations with higher order.

\vspace{1cm}
\noindent

\newpage
\bibliographystyle{plain}
\bibliography{cite}
\newpage

\section{Appendix}
\noindent
The appendix is divided into two parts. We briefly review the
general theory of Chebyshev approximations and
of rational Chebyshev approximations in Section
\ref{Sec:App:ChebyshevApproximations} and
\ref{Sec:App:RationalApproximations}, respectively.
For further background on rational approximations we refer to
Ralston \& Rabinowitz~\cite{Ralston2001} (Chapter 7).

\subsection{Chebyshev Approximations}
\label{Sec:App:ChebyshevApproximations}
\noindent
The Chebyshev polynomial of degree $n$ is defined as
\begin{eqnarray}
T_n(x) = \text{cos} (n \ \text{arc} \ \text{cos} \ x) \ \ \ \text{for} \ -1 \leq x \leq 1,
\label{Eqn:DefinitionTn}
\end{eqnarray}
where the $ T $ has no relation to the $ T $ denoting maturity in Section \ref{Sec:VanillaCallOption}. Even though these terms may look trigonometric, it should be noted that explicit expressions for $ T_n(x)  $ can be found by the following recursion
\begin{eqnarray}
T_0(x) &=& 1 \nonumber \\
T_1(x) &=& x \nonumber \\
T_2(x) &=& 2x^2-1  \label{Eqn:RecursionTn}\\
T_3(x) &=& 4x^3-3x \nonumber \\
&...&  \nonumber\\
T_{n+1}(x) &=& 2x T_{n}(x)-T_{n-1}(x) \ \ \ n \geq 1 \nonumber.
\end{eqnarray}
Each Chebyshev polynomial $ T_n(x) $ is continuous and has $ n $ zeros on the interval $ [-1,1] $. Each of the zeros is located at
\begin{eqnarray}
x = \text{cos} \left( \frac{\pi(k-0.5)}{n} \right) \ \ \ \ \ \  \text{         k = 1,2,...,n},
\end{eqnarray}
whereas the $ n+1 $ extrema are located at
\begin{eqnarray}
x = \text{cos} \left( \frac{\pi k}{n} \right) \ \ \ \ \ \  \text{         k = 1,2,...,n}.
\end{eqnarray}
At its maxima $ T_n(x) = 1 $ and at its minima $ T_n(x) = -1 $ for all $ n $. In addition, the Chebyshev polynomials satisfy a discrete orthogonality relation on $[-1,1]$. If we take an arbitrary function $ f(x) $ with $x \in [-1,1] $ and define
\begin{eqnarray}
\ \ \ \ \ c_j = \frac{2}{N} \sum_{k=1}^{N} f(x_k) T_j(x_k) = \frac{2}{N} \sum_{k=1}^{N} f \left(\text{cos} \left( \frac{\pi(k-0.5)}{N} \right)\right) \text{cos} \left( \frac{\pi j (k-0.5)}{N} \right),
\label{Eqn:ApproximationCj}
\end{eqnarray}
then it can be shown that the approximation
\begin{eqnarray}
f_{Cheb}(x) = \frac{c_0}{2} + \sum_{j=1}^{N-1} c_j T_j(x),
\label{Eqn:RADefinition3}
\end{eqnarray}
is exactly equal to $ f(x) $ at all $ N $ zeros of $ T_n(x) $. Therefore, if we increase $ N $, the two functions will be exact in more and more points. \\
\noindent
If one inserts the expressions given for $ T_n(x) $ in (\ref{Eqn:RecursionTn}) into Equation (\ref{Eqn:RADefinition3}), one obtains a polynomial in $ x $, which approximates the function $ f(x) $ on the interval $[-1,1]$. Even though this Chebyshev approximating polynomial is not equal to the minimax polynomial, which among all polynomials of the same degree has the smallest maximum deviation from the true function, it comes very close. The minimax polynomial is very difficult and time consuming to calculate, whereas the Chebyshev approximating polynomial is almost identical and is extremely easy to compute. \\
\noindent
Finally, if one is interested to approximate a variable $ y  \in [a,b] $, one should use the following change of variable to map it on $[-1,1]$ and proceed as before
\begin{eqnarray}
x = \frac{2y-(b+a)}{b-a}.
\label{Eqn:abxy}
\end{eqnarray}

\subsection{Rational Chebyshev Approximations}
\label{Sec:App:RationalApproximations}
\noindent
In the previous section, we have discussed how to find a good polynomial approximation for a given function $ f(x) $ on a given interval $ [a,b] $. Rather than approximating an arbitrary function $ f(x) $ by a linear combination of Chebyshev polynomials, one can approximate $ f(x) $ by a ratio of linear combinations of these polynomials as follows:
\begin{eqnarray}
f(x) \approx \frac{\sum_{j=0}^m a_j T_j(x)}{\sum_{j=0}^k b_j T_j(x)} = T_{m,k}(x).
\label{Eqn:RADefinition1}
\end{eqnarray}
For simplicity and for all rational function approximations used here, set $ m = k $, and define
\begin{eqnarray}
f(x) \approx \frac{\sum_{j=0}^m a_j T_j(x)}{\sum_{j=0}^m b_j T_j(x)} = f_{RA}^{m} (x),
\label{Eqn:RADefinition2}
\end{eqnarray}
where $ T_n(x) $ are the Chebyshev polynomials defined in (\ref{Eqn:DefinitionTn}). \\
\noindent The coefficients $ a_j $ and $ b_j$ are chosen by assuring as many coefficients of $ T_j(x) $ in the following to be zero
\begin{eqnarray}
f(x) - f_{RA}^{m} (x) = \frac{c_0}{2} + \sum_{j=1}^{\infty} c_j T_j(x) - \frac{\sum_{j=0}^m a_j T_j(x)}{\sum_{j=0}^m b_j T_j(x)}.
\label{Eqn:RADefinition6}
\end{eqnarray}
This equation can be rewritten as
\begin{eqnarray}
\text{                } \ \ \ \ \ \ \ \  f(x) -  f_{RA}^{m} (x) &=& \frac{ \left[ \frac{c_0}{2} + \sum_{j=1}^{\infty} c_j T_j(x) \right] \left[ \sum_{j=0}^m b_j T_j(x) \right] - \sum_{j=0}^m a_j T_j(x)}{\sum_{j=0}^m b_j T_j(x)} \\
&=& \frac{ \frac{c_0}{2} \sum_{j=0}^m b_j T_j(x) + \sum_{j=1}^{\infty} \sum_{i=0}^m \frac{b_j c_j}{2} [T_{i+j}(x)+ T_{|i-j|}(x) ] - \sum_{j=0}^m a_j T_j(x)}{\sum_{j=0}^m b_j T_j(x)}.
\end{eqnarray}
Collecting coefficients of like terms and setting the resulting coefficients to zero, one gets
\begin{eqnarray}
a_0 = \sum_{i=0}^m \frac{b_j c_j}{2},
\label{Eqn:DefinitionA0}
\end{eqnarray}
and
\begin{eqnarray}
a_r = \frac{1}{2} \sum_{i=0}^m b_j (c_{i+r}+ c_{|i-r|}) \ \ \ \ r = 1,2,...,2m+1,
\label{Eqn:DefinitionAr}
\end{eqnarray}
where $a_r=0$ if $ r>m $ and $ b_0 $ is chosen to be $ 1 $ ($ c_j $ is defined as in Equation (\ref{Eqn:ApproximationCj})).  This leads to a linear system of equations for any $ m $, which can be solved easily for the parameters $ a_j $ and $ b_j $ of Equation (\ref{Eqn:RADefinition2}). To illustrate, the following is the resulting system of equations for a $ 4 \times 4 $ rational function approximation:
\begin{eqnarray*}
2 a_0 &=& c_0 + b_1 c_1 + b_2 c_2 + b_3 c_3 + b_4 c_4 \\
2 a_1 &=& 2c_1 + b_1 (c_0 + c_2) + b_2 (c_1 + c_3) + b_3 (c_2 + c_4) + b_4 (c_3 + c_5) \\
2 a_2 &=& 2c_1 + b_1 (c_1 + c_3) + b_2 (c_0 + c_4) + b_3 (c_1 + c_5) + b_4 (c_2 + c_6) \\
2 a_3 &=& 2c_1 + b_1 (c_2 + c_4) + b_2 (c_1 + c_5) + b_3 (c_0 + c_6) + b_4 (c_1 + c_7) \\
2 a_4 &=& 2c_1 + b_1 (c_3 + c_5) + b_2 (c_2 + c_6) + b_3 (c_1 + c_7) + b_4 (c_0 + c_8)  \\
0 &=& 2c_1 + b_1 (c_4 + c_6) + b_2 (c_3 + c_7) + b_3 (c_2 + c_8) + b_4 (c_1 + c_9)  \\
0 &=& 2c_1 + b_1 (c_5 + c_7) + b_2 (c_4 + c_8) + b_3 (c_3 + c_9) + b_4 (c_2 + c_{10}) \\
0 &=& 2c_1 + b_1 (c_6 + c_8) + b_2 (c_5 + c_9) + b_3 (c_4 + c_{10}) + b_4 (c_3 + c_{11}) \\
0 &=& 2c_1 + b_1 (c_7 + c_9) + b_2 (c_6 + c_{10}) + b_3 (c_5 + c_{11}) + b_4 (c_4 + c_{12}).
\end{eqnarray*}
This system of equations can be written in matrix form as follows:
\begin{eqnarray*}
       \ \ \ \ \  \left(
        \begin{array}{cc}
        c_0  \\ c_1 \\ c_2 \\ c_3 \\ c_4\\ c_5 \\ c_6 \\ c_7 \\ c_8
        \end{array}
        \right) = \frac{1}{2} \left(
        \begin{array}{cc}
        -2 \\ 0 \\ 0 \\ 0 \\ 0 \\ 0 \\ 0 \\ 0 \\ 0
        \end{array}
        \begin{array}{cc}
        0 \\  -2 \\ 0 \\ 0 \\ 0 \\ 0 \\ 0 \\ 0 \\ 0
        \end{array}
        \begin{array}{cc}
        0 \\  0 \\ -2 \\ 0 \\ 0 \\ 0 \\ 0 \\ 0 \\ 0
        \end{array}
                \begin{array}{cc}
        0 \\  0 \\ 0 \\ -2 \\ 0 \\ 0 \\ 0 \\ 0 \\ 0
        \end{array}
                \begin{array}{cc}
        0 \\  0 \\ 0 \\ 0 \\ -2 \\ 0 \\ 0 \\ 0 \\ 0
        \end{array}
                \begin{array}{cc}
        c_1 \\ c_2 + c_0 \\ c_3 + c_1 \\ c_4 + c_2 \\ c_5 + c_3 \\ c_6 + c_4 \\ c_7 + c_5 \\ c_8 + c_6 \\ c_9 + c_{7}
        \end{array}
                \begin{array}{cc}
        c_2 \\ c_3 + c_1 \\ c_4 + c_0 \\ c_5 + c_1 \\ c_6 + c_2 \\ c_7 + c_3 \\ c_8 + c_4 \\ c_9 + c_5 \\ c_{10}+ c_{6}
        \end{array}
                \begin{array}{cc}
        c_3 \\ c_4 + c_2 \\ c_5 + c_1 \\ c_6 + c_0 \\ c_7 + c_1 \\ c_8 + c_2 \\ c_9 + c_3 \\ c_{10}+ c_{4} \\ c_{11}+ c_{5}
        \end{array}
                \begin{array}{cc}
        c_4 \\ c_5 + c_3 \\ c_6 + c_2 \\ c_7 + c_1 \\ c_8 + c_0 \\ c_9 + c_1 \\ c_{10}+ c_{2} \\ c_{11}+ c_{3} \\ c_{12}+ c_{4}
        \end{array}
        \right) \left(
        \begin{array}{cc}
                a_0 \\ a_1 \\ a_2 \\ a_3 \\ a_4 \\ b_1 \\ b_2 \\ b_3 \\ b_4
        \end{array}
        \right).
\end{eqnarray*}
Substituting the resulting coefficients of $ a_j $ and $ b_j $ into Equation (\ref{Eqn:RADefinition2}) gives a rational Chebyshev approximation of function $ f(x) $.
\vspace{3cm}

\end{document}